%% file: texturezeros.tex
\documentclass[12pt]{article}
\usepackage{amsmath,amssymb,amsthm}
\usepackage{bbm,latexsym}
\usepackage{graphicx}
\usepackage{rotating} 
\usepackage[numbers,sort&compress]{natbib}
\newtheorem{theorem}{Theorem}
\begin{document}
\title{\normalsize \hfill UWThPh-2014-13 \\[1cm] \LARGE
A complete survey of texture zeros\\ in the lepton mass matrices}
\author{P.O. Ludl\thanks{E-mail: patrick.ludl@univie.ac.at} 
\setcounter{footnote}{6}
and W. Grimus\thanks{E-mail: walter.grimus@univie.ac.at} \\[4mm]
\small University of Vienna, Faculty of Physics \\
\small Boltzmanngasse 5, A--1090 Vienna, Austria \\[4.6mm]}

\date{October 16, 2014}

\maketitle

\begin{abstract}
We perform a systematic and complete analysis of texture zeros in the
lepton mass matrices and identify all viable and maximally restrictive
cases of pairs $(M_\ell, M_D)$ and $(M_\ell, M_L)$, where $M_\ell$,
$M_D$ and $M_L$ are the charged-lepton, Dirac neutrino and Majorana
neutrino mass matrices, respectively.
To this end, we perform a thorough analysis of textures which
are equivalent through weak-basis permutations. Furthermore, we
introduce numerical measures for the predictivity of textures and
apply them to the viable and maximally restrictive texture zero models.
It turns out that for Dirac neutrinos these models can at most predict
the smallest neutrino mass and the CKM-type phase of the mixing matrix.
For Majorana neutrinos most models
can, in addition, predict the effective neutrino mass for neutrinoless
double beta decay.
Apart from one model, which has marginal
predictive power with respect to $\sin^2\! \theta_{23}$,
no other model can predict any of the already
measured observables.
\end{abstract}

\newpage

\section{Introduction}
In spite of the enormous progress in our knowledge of neutrino masses
and lepton mixing~\cite{fits1,fits2,fits3,fogli}, 
the origin of the leptonic flavour structure is
still a mystery. One popular approach to resolve this mystery 
is through underlying symmetries. The simplest symmetries in this
context are Abelian. Such symmetries can be used to impose texture
zeros in the mass matrices in order to make them predictive---see for
instance~\cite{low1,low2} for the lepton sector and~\cite{serodio} for the
quark sector. 
Vice versa, given mass matrices with texture zeros, one can always
find an extended scalar sector and suitable Abelian symmetries such
that the texture zeros originate from these
symmetries~\cite{grimus,gonzalez}. In this sense, texture zeros are synonymous
with Abelian symmetries. In the most simple scenario the 
charged-lepton mass matrix $M_\ell$ is diagonal, which signifies six
texture zeros in $M_\ell$, and, assuming that neutrinos are
Majorana fermions, some texture zeros are placed 
in the Majorana mass matrix $M_L$. 
It has been shown that the data allow seven mass matrices $M_L$ with
two texture zeros~\cite{FGM}. Subsequently, these seven cases have
received a lot of attention---see~\cite{tz1,tz2,tz3,tz4,tz5,tz6,tz7,tz8,tz9,tz10,tz11,tz12,tz13,tz14,tz15,tz16} for an incomplete list
of references. If neutrinos are Dirac particles, then one has more
freedom for texture zeros in the corresponding
mass matrix $M_D$~\cite{Hagedorn-Rodejohann} because $M_D$ is an
arbitrary complex $3 \times 3$ matrix, in contrast to $M_L$ which has
to be symmetric. As for cases with $M_\ell$ non-diagonal, studies are
usually confined to instances where both the charged-lepton and the neutrino
mass matrix have a Fritzsch-like texture~\cite{fritzsch1,fritzsch2} or extensions
of it---see~\cite{fl1,fl2,fl3,fl4,fl5,fl6,fl7,fl8,fl9,fl10,fl11,fl12,fl13,fl14,fl15} and references therein. In general, one can
define ``parallel textures'' as those where $M_\ell$ and $M_L$ or
$M_D$ have the same texture~\cite{weak-basis2,wang}. For a recent review
on textures we refer the reader to~\cite{review}.

In this paper we perform a systematic numerical study of all possibilities of
texture zeros in the charged-lepton and neutrino mass matrices. We
stress that this study also includes all non-parallel
textures. Moreover, we investigate both Dirac and Majorana neutrino
mass matrices and for each of these two neutrino types we
discuss separately normal ordering ($m_1 < m_2 < m_3$) and inverted
ordering ($m_3 < m_1 < m_2$) of the neutrino mass spectrum.
Our main results will be presented as four lists of viable and maximally
restrictive textures, \textit{i.e.}\ those textures with a maximal
number of zeros which are able to reproduce the existing
mass and mixing data. However, it is not only interesting if a texture
is viable, it is also rewarding to know if a texture is
predictive. In order to define such general predictivity
criteria, we note that there are eight known observables
$\mathcal{O}_j$ ($j=1,\ldots,8$) in the lepton sector, 
the three charged-lepton masses, the two neutrino
mass-squared differences and three mixing angles, and five
other observables, on which (almost) nothing is known: the smallest
neutrino mass, the effective neutrino mass in neutrinoless double beta
decay, the CKM-type phase and the two Majorana phases. 
Our predictivity criterion for the eight known observables
will be defined as certain numerical measures
which allow to judge how well the mean value of $\mathcal{O}_i$ can
be predicted from the measured values 
$\overline{\mathcal{O}}_j \pm \sigma_j$ ($j \neq i$) of the seven other
observables. A related but distinct numerical method will be applied
to the five observables which have not yet been measured.
The results of the predictivity analyses will be included in the lists
of viable textures.

The notation of texture zeros in this paper is the usual
one: texture zeros are denoted by
zero entries in the mass matrix, while an entry $\times$ in
a mass matrix denotes an arbitrary complex number. For example,
the matrix
\begin{equation}
\begin{pmatrix}
\times & \times & 0\\
\times & 0      & \times\\
0      & 0      & \times
\end{pmatrix}
\end{equation}
has four texture zeros and the five entries $\times$
denote arbitrary and independent complex numbers. Majorana
mass matrices are understood to be symmetric, so if
\begin{equation}
\begin{pmatrix}
\times & \times & 0\\
\times & 0      & 0\\
0      & 0      & \times
\end{pmatrix}
\end{equation}
represents a Majorana mass matrix, the 11, 12 and 33-entries 
are arbitrary complex numbers, while the 21-element must be equal
to the 12-element. Since this matrix contains only three independent
zero entries, it is counted as an instance of three texture zeros
in a Majorana mass matrix.

The paper is organized as follows. 
In section~\ref{constraints} we discuss basic constraints on the mass
matrices. Section~\ref{weak-basis} is devoted to the relationship between
texture zeros and weak-basis transformations. In
section~\ref{classification} we classify the possible inequivalent
texture-zero models for Dirac and Majorana
neutrinos. Section~\ref{numerical} is the main part of the paper: it
contains the explanation of the numerical analyses, 
the definition of our predictivity measures and the results of our
analyses. Finally, in section~\ref{concl} the conclusions are
presented. Some technical details of the numerics are deferred to an
appendix. 

\section{Basic constraints on the lepton mass matrices}
\label{constraints}
In this paper, one of our basic assumptions is that the lepton masses
and the mixing matrix are obtained, with sufficient accuracy, by the
tree-level mass matrices. 
Thus we require the charged-lepton mass matrix to have rank three,
while the neutrino mass matrix can have rank three or two, since one
neutrino mass being zero is compatible with all experimental data.
For example, the mass matrix
\begin{equation}\label{rank2texture}
\begin{pmatrix}
0 & 0 & 0\\
\times & \times & \times \\
\times & \times & \times
\end{pmatrix}
\end{equation}
will in general have rank two and thus
does not represent an acceptable charged-lepton
mass matrix. However, it represents a
viable Dirac neutrino mass matrix.

In the case of Majorana neutrinos there is another possible
constraint. Namely, the fact that the Majorana mass matrix is
symmetric, in combination with texture zeros, can lead to
matrices with two equal singular values, which we exclude.
An example for such a texture in a Majorana
neutrino mass matrix is
\begin{equation}\label{equalmassestexture}
\begin{pmatrix}
0      & \times & 0\\
\times & 0      & 0\\
0      & 0      & \times
\end{pmatrix}.
\end{equation}

According to this line of reasoning, 
the above phenomenological
requirements on the fermion masses directly exclude some
types of texture zeros in the fermion mass matrices.
In particular, there is a maximal number of texture
zeros in the mass matrices. The fermion mass terms
occurring in this paper and the maximal number of texture
zeros in the mass matrices are listed in table~\ref{massterms}.
Note that in our convention the chiralities in the 
Dirac neutrino mass term
are interchanged with respect to the charged-lepton mass term.
\renewcommand{\arraystretch}{1.4}
\begin{table}
\begin{center}
\begin{tabular}{|l||c|c|c|c|}
\hline
Fermions & mass term & masses & $\text{rank}(M)$ & $n_{0,\text{max}}$ \\
\hline\hline
    Charged leptons &
    $-\overline{\ell_L} M_\ell \ell_R + \mathrm{H.c.}$ &
    $0 < m_e < m_\mu < m_\tau$ &
    $3$ &
    $6$ \\
    \hline
    Dirac neutrinos &
    $-\overline{\nu_R} M_D \nu_L + \mathrm{H.c.}$ &
    \begin{tabular}{@{}l@{}} $0 \leq m_1 < m_2 < m_3$ or \\ $0 \leq m_3 < m_1 < m_2$ \end{tabular} &
    $\geq 2$ &
    $7$ \\
    \hline
    Majorana neutrinos &
    $\dfrac{1}{2} \nu_L^T C^{-1} M_L \nu_L + \text{H.c.}$ &
    \begin{tabular}{@{}l@{}} $0 \leq m_1 < m_2 < m_3$ or \\ $0 \leq m_3 < m_1 < m_2$ \end{tabular} &
    $\geq 2$ &
    $4$ \\
\hline
\end{tabular}
\end{center}
\caption{The mass terms discussed in this paper. Since we consider
  only the Standard Model 
field content (plus three right-handed neutrinos in the case of Dirac
neutrinos), 
all mass matrices are arbitrary complex $3\times 3$-matrices. In
addition, $M_L$ must be symmetric. 
By $\text{rank}(M)$ we denote the required
rank of the mass matrix $M$ and $n_{0,\text{max}}$ is the maximal
number of (independent) 
texture zeros which can be imposed in $M$.}\label{massterms}
\end{table}
\renewcommand{\arraystretch}{1.0}

The mass 
matrices shown in table~\ref{massterms} 
are diagonalized by 
\begin{subequations}
\begin{eqnarray}
U_L^{(\ell)\dagger} M_\ell U_R^{(\ell)} = \mathrm{diag}(m_e,\,m_\mu,\,m_\tau),\\
U_R^{(\nu)\dagger} M_D U_L^{(\nu)} = \mathrm{diag}(m_1,\,m_2,\,m_3),\\
U_L^{(\nu)T} M_L U_L^{(\nu)} = \mathrm{diag}(m_1,\,m_2,\,m_3),
\end{eqnarray}
\end{subequations}
where $U_L^{(\ell)}$, $U_R^{(\ell)}$, $U_L^{(\nu)}$ and $U_R^{(\nu)}$ are
unitary $3\times 3$ matrices. If we impose texture zeros only in $M_\ell$
or only in the neutrino mass matrix, \textit{i.e.}\ $M_D$ in the case of
Dirac and $M_L$ in the case of Majorana neutrinos, only one of the
two matrices $U_L^{(\ell)}$ and $U_L^{(\nu)}$ is restricted.
In this case there are no constraints on the lepton
mixing matrix
\begin{equation}\label{PMNS}
 U_\text{PMNS} = U_L^{(\ell)\dagger} U_L^{(\nu)}.
\end{equation}

\section{Texture zeros and weak-basis transformations}
\label{weak-basis}
\subsection{Physical implications of texture zeros}
\label{section-physical-implications}

In general, we always have the freedom to perform weak-basis
transformations, \textit{i.e.} field redefinitions which leave
the form of the gauge and kinetic terms of the Lagrangian
invariant~\cite{CP-violation,weak-basis1,weak-basis2}.
In this paper we have in mind models with the same fermionic gauge
multiplet structure as the Standard Model.
If we consider an extension of the Standard Model 
by three
right-handed gauge-singlet neutrino fields
$\nu_R$ and assume lepton number conservation, then neutrinos
are Dirac particles and 
the lepton mass terms are given by
\begin{equation}\label{Dirac-nu-massterm2}
-(\overline{\ell_L} M_\ell \ell_R + \overline{\nu_R} M_D \nu_L + \mathrm{H.c.})
\end{equation}
(see table~\ref{massterms}).
Now we perform a field redefinition
\begin{equation}
\ell_{L,R} = V_{L,R}^{(\ell)} \ell_{L,R}',\quad
\nu_{L,R}  = V_{L,R}^{(\nu)}  \nu_{L,R}'.
\end{equation}
In order to leave the kinetic terms invariant,
$V_{L,R}^{(\ell)}$ and $V_{L,R}^{(\nu)}$ must be unitary
and gauge invariance requires $V_L^{(\ell)}=V_L^{(\nu)}\equiv V_L$.
In this new weak basis, the mass terms of
equation~(\ref{Dirac-nu-massterm2}) are given by
\begin{equation}
-(\overline{\ell_L'} M_\ell' \ell_R' + \overline{\nu_R'} M_D' \nu_L' +
\mathrm{H.c.}) 
\end{equation}
with the new mass matrices
\begin{subequations}\label{newmassmatrices}
\begin{equation}\label{weakbasisDirac}
M_\ell'= V_L^\dagger M_\ell V_R^{(\ell)}
\quad \text{and} \quad 
M_D'= V_R^{(\nu)\dagger} M_D V_L.
\end{equation}
A crucial observation in the case of Dirac neutrinos is that 
$V_R^{(\ell)}$ and $V_R^{(\nu)}$ are independent transformation
matrices, which is a consequence of the assumption that the
right-handed fermion fields are gauge singlets.
Obviously, we do not have this freedom in the
case of Majorana neutrinos, 
where the weak-basis transformation on the mass matrices reads
\begin{equation}\label{weakbasisMajorana}
M_\ell'= V_L^\dagger M_\ell V_R^{(\ell)}
\quad \text{and} \quad 
M_L'= V_L^T M_L V_L.
\end{equation}
\end{subequations}
The fermion masses are the singular values of the mass matrices
and are thus invariant under the biunitary
transformation~(\ref{newmassmatrices}).
Clearly, also the lepton mixing matrix $U_\text{PMNS}$ is invariant
under this transformation.

Consequently, whenever two types of texture zeros
are related by a transformation of the
form~(\ref{newmassmatrices}), they lead to the same physical
constraints and are thus physically equivalent. 
This trivial statement has far-reaching consequences: on the one hand
it allows to divide texture zero models into equivalence classes with
the same physical consequences, on the other hand
it relates to the existence of types of texture zeros, which 
\textit{do
not impose any physical constraints}~\cite{weak-basis1,weak-basis2,ludl},
a fact which follows from the following well-known
theorem of linear algebra.
\begin{theorem}\label{QR}
Let $M$ be a complex $n\times n$-matrix. Then there
exist unitary $n\times n$-matrices $W_L$, $W_R$, $\widetilde W_L$,
$\widetilde W_R$ such that
\begin{itemize}
 \item        $W_L^\dagger M$ is an upper triangular matrix,
 \item $\widetilde W_L^\dagger M$ is a  lower triangular matrix,
 \item        $ M W_R$ is an upper triangular matrix,
 \item $ M \widetilde W_R$ is a  lower triangular matrix.
\end{itemize}
These four statements are equivalent to 
the so-called QR, QL, RQ and
LQ-decomposition of complex square matrices,
respectively.
\end{theorem}

Indeed, theorem~\ref{QR} assures us that beginning with \textit{any}
mass matrices $M_\ell$ and $M_D$ (but not $M_L$) we
can always perform a weak-basis transformation~(\ref{newmassmatrices})
such that, in the new basis, the mass matrices are upper or lower triangular
matrices, \textit{i.e.} $M_\ell$ and $M_D$ are of the form
\begin{equation}\label{trivialtextures}
\begin{pmatrix}
\times & \times & \times\\
0      & \times & \times\\
0      & 0      & \times
\end{pmatrix}
\quad \text{or} \quad
\begin{pmatrix}
\times & 0      & 0\\
\times & \times & 0\\
\times & \times & \times
\end{pmatrix}.
\end{equation}
Consequently, any type of texture zeros equivalent to
one of equation~(\ref{trivialtextures}) 
is equivalent to the trivial case of
\textit{no imposed texture zeros at all}
and does not impose any physical
constraints.
The textures
\begin{equation}\label{trivialtextures2}
\begin{split}
& \begin{pmatrix}
\times & \times & \times\\
0      & \times & \times\\
0      & \times  & \times
\end{pmatrix},\,
\begin{pmatrix}
\times & \times & \times\\
0      & \times & \times\\
\times & 0      & \times
\end{pmatrix},\,
\begin{pmatrix}
\times & \times & \times\\
\times & \times & \times\\
0      & 0      & \times
\end{pmatrix},\,
\\
& \begin{pmatrix}
\times & \times & \times\\
0      & \times & \times\\
\times & \times  & \times
\end{pmatrix},\,
\begin{pmatrix}
\times & \times & \times\\
\times & \times & \times\\
0      & \times & \times
\end{pmatrix},\,
\begin{pmatrix}
\times & \times & \times\\
\times & \times & \times\\
\times & 0      & \times
\end{pmatrix},
\\
& \begin{pmatrix}
\times & 0      & 0\\
\times & \times & \times\\
\times & \times & \times
\end{pmatrix},\,
\begin{pmatrix}
\times & 0      & \times\\
\times & \times & 0\\
\times & \times & \times
\end{pmatrix},\,
\begin{pmatrix}
\times & \times & 0\\
\times & \times & 0\\
\times & \times & \times
\end{pmatrix},\,
\\
& \begin{pmatrix}
\times & 0      & \times\\
\times & \times & \times\\
\times & \times & \times
\end{pmatrix},\,
\begin{pmatrix}
\times & \times & 0\\
\times & \times & \times\\
\times & \times & \times
\end{pmatrix},\,
\begin{pmatrix}
\times & \times & \times\\
\times & \times & 0\\
\times & \times & \times
\end{pmatrix}
\end{split}
\end{equation}
are even less restrictive than the ones of
equation~(\ref{trivialtextures}) and thus also
do not impose any constraints on physical observables.

\subsection{Equivalence of texture zeros}

The number of possible cases of texture zeros which can be imposed in
fermion mass matrices is huge,\footnote{Ignoring all phenomenological
  requirements, one would find $2^9 \times 2^9 = 262144$ different
  possibilities for texture zeros 
  in the pair of mass matrices $(M_\ell, M_D)$ and $2^9 \times 2^6 =
  32768$ possible 
  patterns of texture zeros in $(M_\ell, M_L)$.} 
however, as discussed above, not all of them
lead to different physical predictions. Therefore,
it is useful to divide the different patterns of
texture zeros into equivalence classes 
with respect to
weak-basis transformations.
Regarding the arrangement of textures into equivalence
classes, we are only interested in weak-basis transformations
which leave the number of texture zeros in each individual
mass matrix invariant.
In general the only weak-basis transformations
fulfilling this requirement will be the ones where
$V_L$, $V_R^{(\ell)}$ and $V_R^{(\nu)}$
of equation~(\ref{newmassmatrices}) are
of the form $PD$, where
$P$ is one of the six permutation
matrices
\begin{equation}\label{permutationmatrices}
\begin{split}
& P_1  = \begin{pmatrix}
1 & 0 & 0\\
0 & 1 & 0\\
0 & 0 & 1
\end{pmatrix},\quad
P_2 = \begin{pmatrix}
0 & 1 & 0\\
1 & 0 & 0\\
0 & 0 & 1
\end{pmatrix},\quad
P_3 = \begin{pmatrix}
0 & 0 & 1\\
0 & 1 & 0\\
1 & 0 & 0
\end{pmatrix},\\
& P_4 = \begin{pmatrix}
1 & 0 & 0\\
0 & 0 & 1\\
0 & 1 & 0
\end{pmatrix},\quad
P_5 = \begin{pmatrix}
0 & 0 & 1\\
1 & 0 & 0\\
0 & 1 & 0
\end{pmatrix},\quad
P_6 = \begin{pmatrix}
0 & 1 & 0\\
0 & 0 & 1\\
1 & 0 & 0
\end{pmatrix}
\end{split}
\end{equation}
and $D$ is a diagonal phase
matrix~\cite{Hagedorn-Rodejohann}.
Weak-basis transformations where
the three unitary matrices $V_L$, $V_R^{(\ell)}$ and $V_R^{(\nu)}$
are diagonal, leave all texture zeros invariant and can
be used to eliminate unphysical phases in the
elements of the mass matrices. In the numerical studies
carried out for this paper, 
this rephasing freedom is used reduce the number of free parameters.

In sections~\ref{Dirac-neutrinos} and~\ref{Majorana-neutrinos}
we will employ the weak-basis transformations based on the permutation matrices
of equation~(\ref{permutationmatrices}) to divide the possible patterns
of texture zeros in the lepton mass matrices into equivalence classes.

\section{Classification of texture zeros in the lepton mass matrices}
\label{classification}
\subsection{Dirac neutrinos}\label{Dirac-neutrinos}

In this section we will use weak-basis transformation~(\ref{weakbasisDirac})
with $V_L$, $V_R^{(\ell)}$, $V_R^{(\nu)}$ being permutation matrices.
For simplicity, such weak-basis transformations will be called
\emph{weak-basis permutations}.

Our strategy 
for
constructing the 
inequivalent classes
of texture zeros in $M_\ell$ and $M_D$ is
as follows. 
A weak-basis permutation 
can be expressed as a composition of the transformations
\begin{equation}\label{weakbasisMl}
M_\ell \rightarrow V_L^\dagger M_\ell V_R^{(\ell)},
\quad
M_D \rightarrow M_D V_L,
\end{equation}
and
\begin{equation}\label{weakbasisMD}
M_\ell \rightarrow  M_\ell,
\quad
M_D \rightarrow V_R^{(\nu)\dagger} M_D.
\end{equation}
The crucial point in our approach to arrange
the different patterns of texture zeros into classes
is that the two above equations allow us to separately discuss
texture zeros in $M_\ell$ and $M_D$.
In the first step, making use of the weak-basis
transformation~(\ref{weakbasisMl}) 
with $V_L$ and $V_R^{(\ell)}$ being permutation matrices,
we can arbitrarily permute rows and columns of $M_\ell$. Since $V_L$ is
a permutation matrix, the
\textit{set of all patterns of texture zeros in $M_D$} remains invariant
under the transformation $M_D \rightarrow M_D V_L$.
In the second step we use the weak-basis permutation~(\ref{weakbasisMD})
which leaves $M_\ell$ invariant and allows to permute the rows of $M_D$.
In other words, the equivalence classes of texture zeros in $M_\ell$ and $M_D$
may be found as follows:
\begin{enumerate}
    \item We divide the possible patterns of texture zeros in $M_\ell$
      into equivalence classes. 
Two types of texture zeros in $M_\ell$ are equivalent if they can be
transformed into each other 
by permutations of the rows and columns of $M_\ell$.
    \item We divide the possible patterns of texture zeros in $M_D$
      into equivalence classes. 
Two types of texture zeros in $M_D$ are equivalent if they can be
transformed into each other 
by permutations of the rows of $M_D$.
    \item The classes of texture zeros in the pair $(M_\ell, M_D)$ are
      obtained by 
combining the classes of $M_\ell$ with the classes of $M_D$.
\end{enumerate}
Note that following the above prescription, we
do in general \textit{not} exploit the full freedom of weak-basis
permutations. The reason is that for those $M_\ell$ and permutation
matrices $\widetilde{V}_L$, $\widetilde{V}_R^{(\ell)}$ such that 
the transformation
\begin{subequations}\label{remainingfreedom}
\begin{equation}
M_\ell \rightarrow  \widetilde{V}_L^\dagger M_\ell \widetilde{V}_R^{(\ell)}
\end{equation}
leaves the positions of the zeros in $M_\ell$ invariant, we have the freedom 
of multiplying $M_D$ with $\widetilde{V}_L$ 
\textit{from the right}:
\begin{equation}
M_D \rightarrow M_D \widetilde{V}_L.
\end{equation}
\end{subequations}
Thus, in addition to the permutation of rows in
equation~(\ref{weakbasisMD}), there is 
this possibility to permute the columns of $M_D$. 
This means that a further step is required to eliminate the remaining
equivalent classes. We do this by brute force:
\begin{enumerate}
\setcounter{enumi}{3}
\item 
We go through all 
classes of texture zeros in $(M_\ell, M_D)$
found by steps~1, 2 and~3
and perform all possible weak-basis permutations of the
form~(\ref{weakbasisDirac}). By comparison  
we eliminate redundant classes.
\end{enumerate}
At first sight our procedure 
with the four steps 
might look cumbersome, but it has two
important advantages. First, it does not require the explicit 
discussion
of all $2^{18}$ possible textures in $(M_\ell,M_D)$. The other
advantage is that the separate treatment of $M_\ell$
and $M_D$ automatically yields a simple way of labeling the
different textures by combining the label of
$M_\ell$ with the label of $M_D$.

Carrying out steps 1--4, we find the following:
\paragraph{Step 1:} using the requirement that $M_\ell$ must be of
rank three, one finds 247 different patterns of
texture zeros in $M_\ell$. Representatives of each
equivalence class are found by going through the 247 patterns,
removing textures equivalent to an already found one.
Moreover, one has to remove all cases which are equivalent
to one of the textures of equations~(\ref{trivialtextures})
and~(\ref{trivialtextures2}).
Finally, one ends up with only nine classes of
texture zeros in the charged-lepton mass matrix.
Table~\ref{Mltextures} lists one representative of each class.

\begin{table}
\begin{small}
\begin{center}
\include{Mlclasses-table}
\end{center}
\end{small}
\caption{Representatives for the nine classes of texture zeros
in the charged-lepton mass matrix $M_\ell$.}\label{Mltextures}
\end{table}

\paragraph{Step 2:} as discussed before, two types of texture zeros
in $M_D$ are equivalent if they are related
by permutations of the \textit{rows} of $M_D$.
Going through the 478 patterns of texture zeros
of rank at least two, keeping only a representative of each class
and discarding all textures which are equivalent to one of the textures
of equations~(\ref{trivialtextures}) and~(\ref{trivialtextures2}),
we end up with 94 classes of texture zeros in $M_D$. 
We list a representative of each class in tables~\ref{MDtextures1}
and \ref{MDtextures2}.

\begin{table}
\begin{small}
\include{MDclasses-table1}
\end{small}
\caption{Representatives for the 94 classes of texture zeros
in the Dirac neutrino mass matrix $M_D$. Part 1: four and
less texture zeros.}\label{MDtextures1}
\end{table}

\begin{table}
\begin{small}
\include{MDclasses-table2}
\end{small}
\caption{Representatives for the 94 classes of texture zeros
in the Dirac neutrino mass matrix $M_D$. Part 2: five and
more texture zeros.}\label{MDtextures2}
\end{table}

\paragraph{Steps 3 and 4:} up to now, by means of weak-basis permutations, we
have divided the $247 \times 478 = 118066$ possible cases
of texture zeros in $M_\ell$ and $M_D$ into
$9 \times 94 = 846$ classes. All members of a class
make the same physical predictions. A representative
of each class is obtained by combining a texture of
$M_\ell$, \textit{cf.}\ table~\ref{Mltextures}, with one of $M_D$,
\textit{cf.}\ tables~\ref{MDtextures1} and~\ref{MDtextures2}.
However, as discussed before, some of the 846 classes are redundant.
By applying all $6^3=216$ weak-basis permutations 
to the 846 classes, we find that 276 of them are redundant, leaving
a total of 570 non-equivalent classes of texture zeros in $(M_\ell,M_D)$.
We show the redundant classes in table~\ref{Ml-MD-redundant}.
\begin{table}
\begin{center}
\begin{small}
\include{Ml-MD-classes_redundant}
\end{small}
\caption{The 276 redundant classes of texture zeros in the pair $(M_\ell,M_D)$.}\label{Ml-MD-redundant}
\end{center}
\end{table}

Finally, we want to emphasize an important issue concerning our treatment
of texture zeros. This issue becomes most striking in comparison with
the treatment of texture zeros in the literature
when $M_\ell$ is diagonal and assumed to be
\begin{equation}\label{diagMl}
	M_\ell = \mathrm{diag}(m_e,\, m_\mu,\, m_\tau).
\end{equation}
This charged-lepton mass matrix corresponds to our texture
\begin{equation}
M_\ell \sim 6_1^{(\ell)} \sim
  \begin{pmatrix}
  0 & 0 & \times \\
  0 & \times & 0 \\
  \times & 0 & 0
  \end{pmatrix}
\sim
  \begin{pmatrix}
  \times & 0 & 0 \\
  0 & \times & 0 \\
  0 & 0 & \times
  \end{pmatrix}.
\label{our-diagMl}
\end{equation}
However, with our numerical procedure we cannot fix the order in which 
the charged-lepton masses $m_e$, $m_\mu$, $m_\tau$ appear on the
diagonal of $M_\ell$ and $U_L^{(\ell)}$ will in general be 
a permutation matrix.
Therefore, the classification of texture zeros in $M_D$
of Hagedorn and Rodejohann in~\cite{Hagedorn-Rodejohann}, 
which assumes the validity of equation~(\ref{diagMl}) and which has,
therefore, always $U_L^{(\ell)} = \mathbbm{1}$,
has no unique correspondence to our notation. However,
one may view the textures studied in~\cite{Hagedorn-Rodejohann}
as special representatives of the classes of texture zeros
shown in table~\ref{correspondenceHR}.

\renewcommand{\arraystretch}{1.4}
\begin{table}
\begin{center}
	\begin{tabular}{|lc|lc|lc|lc|}
	\hline
	$A$ & $6_1^{(\ell)}-5_6^{(\nu_D)}$ & $\widetilde{D}_1$ & $6_1^{(\ell)}-4_{16}^{(\nu_D)}$ & $F_3$ & $6_1^{(\ell)}-4_{3}^{(\nu_D)}$ & $G_7$ & $6_1^{(\ell)}-3_{6}^{(\nu_D)}$ \\
	$B$ & $6_1^{(\ell)}-5_4^{(\nu_D)}$ & $\widetilde{D}_2$ & $6_1^{(\ell)}-4_{8}^{(\nu_D)}$ & $G_1$ & --- & $G_8$ & $6_1^{(\ell)}-3_{14}^{(\nu_D)}$ \\
	$\widetilde{B}$ & $6_1^{(\ell)}-5_5^{(\nu_D)}$ & $\widetilde{D}_3$ & $6_1^{(\ell)}-4_{23}^{(\nu_D)}$ & $G_2$ & $6_1^{(\ell)}-3_{4}^{(\nu_D)}$ & $G_9$ &---\\
	$C$ & $6_1^{(\ell)}-4_{24}^{(\nu_D)}$ & $E$ & $6_1^{(\ell)}-4_{9}^{(\nu_D)}$ & $G_3$ & $6_1^{(\ell)}-3_{2}^{(\nu_D)}$ & $G_{10}$ & $6_1^{(\ell)}-3_{12}^{(\nu_D)}$\\
	$D_1$ & $6_1^{(\ell)}-4_7^{(\nu_D)}$ & $\widetilde{E}$ & $6_1^{(\ell)}-4_{17}^{(\nu_D)}$ & $G_4$ & $6_1^{(\ell)}-3_{3}^{(\nu_D)}$ & & \\
	$D_2$ & $6_1^{(\ell)}-4_{15}^{(\nu_D)}$ & $F_1$ & $6_1^{(\ell)}-4_{1}^{(\nu_D)}$ & $G_5$ & $6_1^{(\ell)}-3_{5}^{(\nu_D)}$ & & \\
	$D_3$ & $6_1^{(\ell)}-4_{22}^{(\nu_D)}$ & $F_2$ & $6_1^{(\ell)}-4_{2}^{(\nu_D)}$ & $G_6$ & $6_1^{(\ell)}-3_{10}^{(\nu_D)}$ & & \\
	\hline
	\end{tabular}
\end{center}
\caption{The types of texture zeros in the Dirac neutrino mass matrix, assuming a diagonal
charged-lepton mass matrix studied by Hagedorn and Rodejohann in~\cite{Hagedorn-Rodejohann}.
Left: notation of~\cite{Hagedorn-Rodejohann}, right: corresponding class of textures in our notation.
$G_1$ and $G_9$ of~\cite{Hagedorn-Rodejohann} have no correspondence in our paper,
because these textures do not imply any physical constraints---see the discussion at the
end of section~\ref{section-physical-implications}.}\label{correspondenceHR}
\end{table}
\renewcommand{\arraystretch}{1.0}

\subsection{Majorana neutrinos}\label{Majorana-neutrinos}

In the case of Majorana neutrinos, the discussion
of the possible patterns of texture
zeros in $M_\ell$ is the same as in
section~\ref{Dirac-neutrinos}. However, under
weak-basis permutations the Majorana neutrino mass
matrix transforms as
\begin{equation}
M_L \rightarrow M_L' = V_L^T M_L V_L,
\end{equation}
\textit{i.e.}\ 
there is no unitary transformation $V_R^{(\nu)}$.
Thus, in contrast to the case of $M_D$, after having
already ``used up'' the freedom of choosing $V_L$ when
dividing the textures of $M_\ell$ into classes, there is
no freedom left to divide the textures of $M_L$. 
Consequently, we have to investigate \textit{all} possible
texture zeros in $M_L$, taking into account that $M_L$ is
symmetric.
Note that, since now there is no freedom of performing
weak-basis transformations, also theorem~\ref{QR} does
not apply to $M_L$. However, the trivial case of no texture zeros
at all still has to be excluded. Thus, $M_L$ must have at least
one and at most four texture zeros---see table~\ref{massterms}.
Furthermore, some textures with four zeros lead to
two degenerate neutrino masses, which is phenomenologically
excluded.
Going through all possible patterns of texture zeros in $M_L$,
keeping only those which are of rank at least two and fulfill the
above requirements, we find 50 possible textures, which are listed
in table~\ref{MnuLtextures}.
Thus in total we find $9 \times 50 = 450$ types of
texture zeros in $M_\ell$ and $M_L$.
As in the case of
Dirac neutrinos, by separately treating $M_\ell$ and $M_L$, we have ignored
up to now the possibility that a weak-basis permutation with
$\widetilde{V}_L$ and $\widetilde{V}_R^{(\ell)}$ leaves the
positions of the zeros in $M_\ell$ invariant, in which case the transformation
\begin{equation}\label{remainingfreedomMajorana}
M_\ell \rightarrow  \widetilde{V}_L^\dagger M_\ell \widetilde{V}_R^{(\ell)},
\quad
M_L \rightarrow \widetilde{V}_L^T M_L \widetilde{V}_L
\end{equation}
is allowed.
Therefore, we have to go through all possible weak-basis permutations
to eliminate the redundant classes.
Doing so, we find 152 classes to be redundant, leaving a total
of 298 classes of texture zeros in $(M_\ell,M_L)$.
The redundant classes are presented in table~\ref{ML-redundant}.

Analogous to the case of Dirac neutrinos---see discussion at 
the end of section~\ref{Dirac-neutrinos}---the 
textures with $M_\ell \sim 6_1^{(\ell)}$ have \textit{no one-to-one
correspondence} with the textures with
$M_\ell = \mathrm{diag}(m_e,\,m_\mu,\,m_\tau)$ studied in~\cite{FGM}. 
Indeed the seven types of
two texture zeros of~\cite{FGM} are special cases of
classes of texture zeros discussed in the present paper:
$A_1$, $A_2$, $B_3$ and $B_4$ belong to the
same class $6_1^{(\ell)}-2_4^{(\nu_L)}$, the textures $B_1$ and $B_2$
are both contained in $6_1^{(\ell)}-2_6^{(\nu_L)}$, and $C$ is a
special case of $6_1^{(\ell)}-2_1^{(\nu_L)}$.

\begin{table}
\begin{small}
\include{MnuLclasses-table}
\end{small}
\caption{Representatives for the 50 classes of texture zeros
in the Majorana neutrino mass matrix $M_L$.}\label{MnuLtextures}
\end{table}

\begin{table}
\begin{center}
\begin{small}
\include{Ml-MnuL-classes_redundant}
\end{small}
\caption{The 152 redundant classes of texture zeros in $(M_\ell, M_L)$.}\label{ML-redundant}
\end{center}
\end{table}

\subsection{The family tree of texture zeros}
\label{family-tree}

\begin{figure}[htb]
\centerline{%
\includegraphics[width=0.6\textwidth,angle=0]{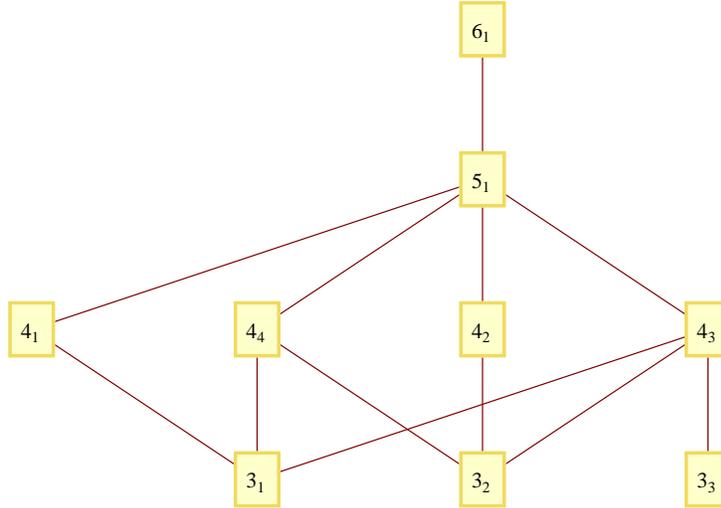}}
\caption{The family tree of texture zeros in $M_\ell$.}\label{Ml-tree}
\end{figure}

If a set of texture zeros in $(M_\ell,M_D)$ or $(M_\ell,M_L)$ 
is compatible with the experimental data, then 
any pattern of texture zeros with one
or more zeros being replaced by free parameters will also
be compatible with the data.
Thus it is sufficient to discuss only those
textures compatible with the experimental data which
are~\textit{maximally restrictive}. By maximally restrictive we
mean that one cannot place a further texture zero 
into one of the two mass matrices
while keeping the model compatible with the data.
For illustrational purposes we can arrange the textures in
a ``family tree,'' where less restrictive textures are the
``children'' of the more restrictive ones. The family tree of
texture zeros in $M_\ell$ is shown in figure~\ref{Ml-tree}.
By using the family trees for texture zeros in the pairs
$(M_\ell,M_D)$ and $(M_\ell,M_L)$, we can easily find the
maximally restrictive patterns.\footnote{The family trees for texture zeros
in $M_D$ and $M_L$ as well as the combinations of the charged-lepton
mass matrix with the neutrino 
mass matrices are too large to be printed here.} 
The list of all allowed patterns
of texture zeros can then be obtained by removing zeros from the maximally
restrictive textures. In our results, tables~\ref{Ml-MD-normal} 
to~\ref{Ml-MnuL-inverted}, we present only the maximally restrictive
pairs of charged-lepton and neutrino mass matrices. 

\section{Numerical analysis}
\label{numerical}
\subsection{$\chi^2$-analysis}\label{chisquared-analysis}

We perform a $\chi^2$-analysis
of the different patterns of texture zeros. Our $\chi^2$-function 
has the usual form
\begin{equation}\label{def-chisquared}
\chi^2(x)\equiv \sum_{i}
\left(
\frac{P_i(x)-\overline{\mathcal{O}}_i}{\sigma_i}
\right)^2,
\end{equation}
where the vector $x$ contains the model parameters, $P_i(x)$ is the
model prediction 
for the observable $\mathcal{O}_i$ and $\overline{\mathcal{O}}_i$ is
the central value 
for $\mathcal{O}_i$. The $\sigma_i$ are the errors of $\mathcal{O}_i$,
where in case of 
asymmetric error distributions we use $\sigma_i^\text{left}$ and
$\sigma_i^\text{right}$ 
for $P_i(x) \leq \overline{\mathcal{O}}_i$ and 
$P_i(x) > \overline{\mathcal{O}}_i$, 
respectively. In this paper we fit the lepton mass matrices with
texture zeros to the eight observables
\begin{displaymath}
	m_e,\, m_\mu,\, m_\tau,\, \Delta m_{21}^2,\, \Delta m_{31}^2,\,
	\mathrm{sin}^2\theta_{12},\, \mathrm{sin}^2\theta_{23}
	\quad\text{and}\quad \mathrm{sin}^2\theta_{13}.
\end{displaymath}
We do not fit the Dirac phase $\delta$. 
The central
values and errors 
of the five neutrino oscillation parameters are taken from the global
fit of oscillation 
data by Fogli~\textit{et al.}~\cite{fogli}. As central values of the
charged-lepton masses we take the 
values from the 
Review of Particle Physics~\cite{PDG}. 
Since the experimental errors on 
the charged-lepton masses are so small that they can cause problems in
the numerical 
$\chi^2$-analysis, we set them to one percent,
\textit{i.e.}\ $\sigma(m_\ell) = 0.01 m_\ell$ for $\ell =
e,\mu,\tau$. For details 
of the numerical implementation of the $\chi^2$-minimization, we refer
the reader to appendix~\ref{details-numerics}.

\paragraph{Results of the $\chi^2$-analysis:} the only information
we can obtain from the standard $\chi^2$-analysis is whether a given
set of texture zeros in the mass matrices is excluded by the experimental
data or not. We use the following criterion for a texture being
compatible with the data:
\begin{quote}
We call a set of textures in the lepton mass matrices
\textit{compatible} with the data if at the minimum of $\chi^2$
the contribution of each observable to $\chi^2$ is 
at most $25$,
\textit{i.e.}\ the deviation of the observable from its
experimental value is at most $5\sigma$.
This implies that $\chi^2_\text{min}\leq 200$.
\end{quote}
We find that according to this criterion about three quarters of 
the classes of texture zeros comprise viable textures.
In the case of Dirac neutrinos all textures
fulfilling the criterion have $\chi^2_\text{min} < 10^{-4}$.
In the case of Majorana neutrinos about 90 percent of the textures
have $\chi^2_\text{min} < 10^{-4}$, the remaining 10 percent 
have
$\chi^2_\text{min} < 30$ (normal spectrum) and $\chi^2_\text{min} < 1$
(inverted spectrum).

Using the ``family tree'' of
texture zeros discussed in section~\ref{family-tree}, we can
identify 
the maximally restrictive textures 
among the compatible ones.
This reduces the number of models to be investigated further to
about 30 each for Dirac and Majorana neutrinos and for both
neutrino mass spectra---see table~\ref{chisquared-results}.
The compatible and maximally restrictive models are presented
in tables~\ref{Ml-MD-normal} to~\ref{Ml-MnuL-inverted}.
In only about a tenth of the compatible and maximally restrictive
textures the charged-lepton mass matrix is diagonal,
\textit{i.e.}\ $M_\ell \sim 6_1^{(\ell)}$.

To summarize, by means of a $\chi^2$-analysis and the
``family tree'' of texture zeros we have 
arrived at
the list of maximally restrictive 
texture zero models
compatible with
the experimental or observational data.
In the following we will further investigate 
each model with regard to its predictivity.

\renewcommand{\arraystretch}{1.4}
\begin{table}
\begin{center}
\begin{tabular}{|l||c|c|c|c|}
\hline
Neutrino nature & \multicolumn{2}{c|}{Dirac} & \multicolumn{2}{c|}{Majorana}\\
\hline
Neutrino mass spectrum & normal & inverted & normal & inverted \\
\hline
Number of textures & $570$ & $570$ & $298$ & $298$ \\
\hline
Compatible with experiment & $430$ & $429$ & $215$ & $229$ \\
\hline
Compatible and maximally restrictive & $29$ & $28$ & $27$ & $25$ \\
\hline
\end{tabular}
\end{center}
\caption{Results of the $\chi^2$-analysis.}\label{chisquared-results}
\end{table}
\renewcommand{\arraystretch}{1.0}

\subsection{Predictivity analysis}
\label{predictivity}
The $\chi^2$-analysis of the previous section only
tells us which textures are compatible with the
observations, but does not yield any statement
on their predictive power.
Therefore, we 
have developed 
a numerical method to estimate
the predictive power of texture zeros in the lepton
mass matrices. The main idea behind this method is
to find an answer to the question:
\begin{quote}
        Given matrices with a viable set of texture zeros and
	fixing the observables
	$\mathcal{O}_j$ $(j\neq i)$ to their experimentally
	observed values, how much can the remaining
	observable $\mathcal{O}_i$ at most deviate from 
	its experimental or best-fit value?
\end{quote}
In the following, we will outline our attempt to
answer the above question for the observables we are
interested in. 
Technical details of the numerical implementation of the
method are presented in appendix~\ref{details-numerics}.

Consider a model with parameters $x$ making predictions
$P_i(x)$ for the observables $\mathcal{O}_i$ with mean
values $\overline{\mathcal{O}}_i$ and errors $\sigma_i$.
For each observable $\mathcal{O}_i$, 
$\chi^2(x)$
has the contribution\footnote{As described
in section~\ref{chisquared-analysis}, in the case of asymmetric error intervals
we use $\sigma_i^\text{left}$ and $\sigma_i^\text{left}$ instead of
a single $\sigma_i$.}
\begin{equation}
\chi_i^2(x)\equiv 
\left(
\frac{P_i(x)-\overline{\mathcal{O}}_i}{\sigma_i}
\right)^2.
\end{equation}
The contributions of all other observables to $\chi^2(x)$
are then given by
\begin{equation}
\widetilde\chi_{i}^2(x)\equiv \sum_{j\neq i}
\left(
\frac{P_j(x)-\overline{\mathcal{O}}_j}{\sigma_j}
\right)^2
= \chi^2(x) - \chi_i^2(x).
\end{equation}
We define a measure for the maximal deviation of
the observable $\mathcal{O}_i$ from its experimentally
observed value as
\begin{equation}\label{predictivity-measure}
	\Delta(\mathcal{O}_i) \equiv \max_{x \in B_i} \chi_i^2(x),
\end{equation}
where $B_i$ is defined as
\begin{equation}\label{def-B_i}
	B_i \equiv
	\left\{ \; x \; \vert \enspace
	\widetilde{\chi}_i^2(x) \leq \chi^2_\text{min} + \delta\chi^2
	\quad \text{and} \quad
	\chi_j^2(x) \leq 25 \enspace \forall j\neq i
	\right\}
\end{equation}
and $\chi^2_\text{min}$ is the minimum of $\chi^2(x)$ found
in the $\chi^2$-analysis of section~\ref{chisquared-analysis}.
The condition $x \in B_i$ fixes the other observables $\mathcal{O}_j$
$(j\neq i)$ to be close to their observed
values $\overline{\mathcal{O}}_j$.\footnote{This is done through the
requirement $\widetilde{\chi}_i^2(x) \leq \chi^2_\text{min} + \delta\chi^2$.
In addition, by the second requirement we demand that 
no observable $\mathcal{O}_j$
is allowed to deviate from 
its central value by more than $5\sigma$.}
The term $\delta\chi^2$ is added to $\chi^2_\text{min}$ in order
to improve convergence of the numerical maximization of $\chi_i^2(x)$
in equation~(\ref{predictivity-measure}).
In this paper, depending on the observable, we use either
$\delta\chi^2=0$ or $\delta\chi^2=1$---see appendix~\ref{details-numerics}.
The quantity $\Delta$ defined in equation~(\ref{predictivity-measure})
allows us to estimate the power of the studied set of texture zeros
to predict $\mathcal{O}_i$. We will use this measure for a
``predictivity analysis'' of the
five neutrino oscillation parameters and define:
\begin{quote}
A set 
of texture zeros can correctly predict
the observable $\mathcal{O}_i$, where  
$\mathcal{O}_i$ is one of the observables 
$\Delta m_{21}^2,\, \Delta m_{31}^2,\, \mathrm{sin}^2\theta_{12},\,
\mathrm{sin}^2\theta_{23},\, \mathrm{sin}^2\theta_{13}$,
if
\end{quote}
\begin{equation}\label{condition-oscill-param}
\Delta(\mathcal{O}_i) \leq 100.
\end{equation}
In other words, we stipulate that a set of texture zeros
is capable to predict an observable $\mathcal{O}_i$ if 
its value can deviate from its central value $\overline{\mathcal{O}}_i$
by at most $10\sigma$, while the 
other observables $\mathcal{O}_j$ $(j\neq i)$ are kept close to their
experimental or best-fit values.

For the charged-lepton masses we use a different
predictivity measure. Namely, for each charged-lepton
mass $m_\ell$ $(\ell = e,\mu,\tau)$ we compute its minimal
and maximal values
\begin{equation}\label{optim-ml}
	m_\ell^\mathrm{min} \equiv \min_{x \in B_{m_\ell}} m_\ell(x)
	\quad \text{and} \quad
	m_\ell^\mathrm{max} \equiv \max_{x \in B_{m_\ell}} m_\ell(x)
\end{equation}
and define:
\begin{quote}
A set 
of texture zeros can correctly predict
the charged-lepton mass $m_\ell$ $(\ell = e,\mu,\tau)$ if
\end{quote}
\begin{equation}\label{criterion-ml}
	m_\ell^\mathrm{min} > \frac{1}{2} m_\ell^\mathrm{exp}
	\quad \text{and} \quad
	m_\ell^\mathrm{max} < 2 m_\ell^\mathrm{exp}.
\end{equation}
Here $m_\ell^\mathrm{exp}$ denotes the mean experimental value
of the charged-lepton mass $m_\ell$ taken from~\cite{PDG}.
In words, we call a model predictive if it predicts
that $m_\ell$
lies between $m_\ell^\mathrm{exp}/2$ and $2 m_\ell^\mathrm{exp}$.

Finally, we also want to define a predictivity measure for those
observables $\mathcal{O}$
which have not been measured up to now. In this case we
compute the minimal and maximal value of 
$\mathcal{O}$
as
\begin{equation}\label{optim-O}
	\mathcal{O}^\mathrm{min} \equiv \min_{x \in B} \mathcal{O}(x)
	\quad \text{and} \quad
	\mathcal{O}^\mathrm{max} \equiv \max_{x \in B} \mathcal{O}(x),
\end{equation}
where the parameter set $B$ is defined as
\begin{equation}
	B \equiv
	\left\{ \; x \; \vert \enspace
	\chi^2(x) \leq \chi^2_\text{min} + \delta\chi^2
	\quad \text{and} \quad
	\chi_j^2(x) \leq 25 \enspace \forall j
	\right\}.
\end{equation}
Thus $B$ is similar to $B_i$ with $\widetilde{\chi}_i^2(x)$ replaced
by the full $\chi^2$-function $\chi^2(x)$. In the numerical analysis
performed
for this paper 
we have computed $\mathcal{O}^\mathrm{min}$ and $\mathcal{O}^\mathrm{max}$
for the observables
\begin{equation}
	m_0,\, m_{\beta\beta},\, \delta,\, \rho \quad\text{and}\quad \sigma.
\end{equation}
Here $m_0$ denotes the mass of the lightest neutrino, \textit{i.e.}\ 
$m_0 = m_1$
in the case of a normal and 
$m_0 = m_3$ 
in the case of an inverted neutrino mass
spectrum.
The effective neutrino mass for neutrinoless double beta decay
is given by
\begin{equation}
m_{\beta\beta} =
\left\vert \sum_{k=1}^3 (U_{\mathrm{PMNS}})_{ek}^2 m_k \right\vert.
\end{equation}
The Dirac CP-phase $\delta$ and the two
Majorana phases $\rho$ and $\sigma$ are defined via the decomposition
\begin{equation}
	U_\text{PMNS} = \exp(i\,  \mathrm{diag}(\alpha,\, \beta,\,
        \gamma)) \times  
	\mathcal{U}(\theta_{12},\theta_{23},\theta_{13},\delta) \times
        \exp(i\,  \mathrm{diag}(\rho,\, \sigma,\, 0)),
\end{equation}
where $\mathcal{U}$ stands for the standard parameterization of the
mixing matrix~\cite{PDG}. 
The phases $\alpha$, $\beta$
and $\gamma$ are not accessible by experimental scrutiny.
By definition, the range of the phases $\delta$, $\rho$ and $\sigma$
is $[0,2\pi)$. 
Since in all our numerical investigations we impose the constraint~\cite{PDG}
\begin{equation}\label{cosmo}
	\sum_{k=1}^3 m_k < 1\,\text{eV}
\end{equation}
on the absolute neutrino mass scale, $m_0$ and $m_{\beta\beta}$ can
assume values between zero and about 
$1/3~\text{eV}$. Given these bounds, we may define:
\begin{quote}
A set 
of textures zeros can predict one of the observables
$m_0$, $m_{\beta\beta}$, $\delta$, $\rho$, $\sigma$ if
\end{quote}
\begin{equation}\label{condition-other-obs}
	\frac{\mathcal{O}^\mathrm{max} - \mathcal{O}^\mathrm{min}}{\mathrm{range}(\mathcal{O})} \leq 0.2.
\end{equation}
Here $\mathrm{range}(m_0) = \mathrm{range}(m_{\beta\beta})= 1/3~\text{eV}$ and
$\mathrm{range}(\delta)=\mathrm{range}(\rho)=\mathrm{range}(\sigma)=2\pi$.

\paragraph{Results of the predictivity analysis:} 
We have performed the analysis explained above
for \emph{all} viable and maximally restrictive texture-zero models.
The results of this paper, \textit{i.e.}\ the viable and maximally
restrictive textures and their predictions, 
are presented in four tables:
\begin{enumerate}
\renewcommand{\labelenumi}{\roman{enumi}.}
\item
table~\ref{Ml-MD-normal}: Dirac neutrinos with normal ordering of the
neutrino masses,
\item
table~\ref{Ml-MD-inverted}: Dirac neutrinos with inverted ordering of the
neutrino masses,
\item
table~\ref{Ml-MnuL-normal}: Majorana neutrinos with normal ordering of the
neutrino masses,
\item
table~\ref{Ml-MnuL-inverted}: Majorana neutrinos with inverted ordering of the
neutrino masses.
\end{enumerate}
In these tables, $n$ denotes the number of real parameters
of the model 
after removing as many phases as possible from the elements of the mass matrices
by means of weak-basis transformations.

The 
results of the predictivity analysis may be summarized as follows.
\begin{itemize}
\item 
According to the criterion of equation~(\ref{criterion-ml}),
none of the investigated textures 
can predict any of the charged-lepton masses. Also relaxing the
condition~(\ref{criterion-ml}) for $m_e$ 
to $m_e^\mathrm{min} > 0.1~\text{MeV}, m_e^\mathrm{max} <
5~\text{MeV}$ does not change this result. 
\item 
Only one set of texture zeros discussed in this paper fulfills the
requirement of equation~(\ref{condition-oscill-param}). Namely, for
the texture $(M_\ell, M_L) \sim (6_1^{(\ell)}, 2_1^{(\nu_L)})$
with an inverted neutrino mass spectrum one has
$\Delta(\mathrm{sin}^2\theta_{23}) = 53.8 <100$.
However, tightening the criterion of equation~(\ref{condition-oscill-param}) to
$\Delta(\mathcal{O}_i)\leq 25$, \textit{i.e.}\ to $5\sigma$, there is no texture
which can predict any of the charged-lepton masses or the neutrino oscillation parameters.
\item 
Most of the investigated textures can predict the smallest
neutrino mass $m_0$.
In the case of Dirac neutrinos, all but
two of the maximally restrictive viable textures have
$\mathrm{rank}(M_D)=2$ and thus $m_0=0$.
\item 
For all but one of the maximally predictive
and compatible textures for Dirac neutrinos we find
$\delta^\mathrm{min} \approx 0$ and $\delta^\mathrm{max} \approx 2\pi$.
The only exception is the texture $(M_\ell, M_D) \sim (6_1^{(\ell)}, 4_5^{(\nu_D)})$
with a normal neutrino mass spectrum, where one finds
$\delta^\mathrm{min} = 0$ and $\delta^\mathrm{max} = \pi$ within the
numerical accuracy.
In fact, through a weak-basis transformation~(\ref{weakbasisDirac})
with $V_L$, $V_R^{(\ell)}$ 
and $V_R^{(\nu)}$ being diagonal phase matrices, in this case
$M_\ell$ and $M_D$ can be made real \textit{simultaneously}, which
implies $\delta\in\{0,\pi\}$. Therefore, this texture does not admit CP
violation in neutrino oscillations.
\item
Many of the textures for the Majorana neutrino
case are predictive with respect to the Dirac phase $\delta$.
In contrast to the case of Dirac neutrinos, here also $\delta \neq 0,\pi$
is possible.
\item 
None of the textures for Majorana neutrinos can predict any of the
Majorana phases $\rho$ and $\sigma$ according to the condition of
equation~(\ref{condition-other-obs}). 
\item 
Almost all maximally restrictive and compatible sets of textures in
$(M_\ell, M_L)$ predict $m_{\beta\beta}$.
\item
The effective mass $m_{\beta\beta}$ can be big (larger than 0.1\,eV) or
small for normal ordering of the neutrino mass spectrum, however, for
the inverted spectrum we always find $m_{\beta\beta} < 0.1$\,eV.
\item
There are a few instances where $M_D$ or $M_L$ are diagonal and lepton
mixing comes purely from $M_\ell$.
Since we deal with maximally restrictive textures, in all of these
instances we trivially have $m_0 = 0$.
\end{itemize}
Thus, the most interesting results of the predictivity analysis
are the minima and maxima of $m_0$, $\delta$ and $m_{\beta\beta}$,
which we also show in tables~\ref{Ml-MD-normal}
to~\ref{Ml-MnuL-inverted}.
As for $\cos\delta$ and Majorana textures with diagonal $M_\ell$---see
the last three lines
in tables~\ref{Ml-MnuL-normal}
and~\ref{Ml-MnuL-inverted}, we have checked that our results agree
with those of~\cite{tz8}.

\paragraph{A note of caution on the interpretation of the results:} 
The presented method 
for estimating the predictivity of a texture-zero model is
based 
on \textit{maximal deviations from the
observed value}, see the definition of $\Delta$ in
equation~(\ref{predictivity-measure}), 
and \textit{maximal and minimal values of observables}, see
equations~(\ref{optim-ml}) and~(\ref{optim-O}). Therefore, 
some of the models
may still possess predictive power 
which can, however, not be
measured by these quantities. 
For example, the texture $(M_\ell, M_D) \sim (6_1^{(\ell)}, 4_5^{(\nu_D)})$
allows $\delta$ to assume
only the values zero and $\pi$, \textit{i.e.}\ we would call this
model ``predictive'' 
with respect to $\delta$.
But in this case our analysis
does not detect
predictive power because equation~(\ref{condition-other-obs}) 
then reads
\begin{equation}
\frac{\delta^\mathrm{max} - \delta^\mathrm{min}}{\mathrm{range}(\delta)}
=
\frac{\pi - 0}{2\pi} = \frac{1}{2} > 0.2.
\end{equation}

\begin{table}
\begin{center}
\begin{small}
\include{Ml-MD-normal}
\end{small}
\caption{The maximally restrictive and compatible classes of
texture zeros in $(M_\ell,M_D)$.
The number of physical parameters
of the texture is denoted by $n$.
For all textures in this table $\chi^2_\mathrm{min}<10^{-4}$. For the
Dirac phase $\delta$ one finds $\delta^\mathrm{min}\approx 0$ and $\delta^\mathrm{max}\approx 2\pi$
for all textures except $6_1-4_5$, where $\delta = 0$ or $\pi$.
Part 1: normal neutrino mass spectrum.}\label{Ml-MD-normal}
\end{center}
\end{table}

\begin{table}
\begin{center}
\begin{small}
\include{Ml-MD-inverted}
\end{small}
\caption{The maximally restrictive and compatible classes of
texture zeros in $(M_\ell,M_D)$. The number of physical parameters
of the texture is denoted by $n$. For all textures in this table
$\chi^2_\mathrm{min}<10^{-4}$, $\delta^\mathrm{min}\approx 0$ and $\delta^\mathrm{max}\approx 2\pi$.
Part 2: inverted neutrino mass spectrum.}\label{Ml-MD-inverted}
\end{center}
\end{table}

\begin{sidewaystable}
\begin{small}
\include{Ml-MnuL-normal}
\end{small}
\caption{The maximally restrictive and compatible classes of
texture zeros in $(M_\ell,M_L)$. The number of physical parameters
of the texture is denoted by $n$. Part 1: normal neutrino mass spectrum.}\label{Ml-MnuL-normal}
\end{sidewaystable}

\begin{sidewaystable}
\begin{small}
\include{Ml-MnuL-inverted}
\end{small}
\caption{The maximally restrictive and compatible classes of
texture zeros in $(M_\ell,M_L)$. The number of physical parameters
of the texture is denoted by $n$. Part 2: inverted neutrino mass spectrum.}\label{Ml-MnuL-inverted}
\end{sidewaystable}

\section{Conclusions}
\label{concl}
In this paper we have performed a
thorough analysis
of texture zeros in
lepton mass matrices for both Dirac and Majorana neutrinos.
It is instructive to summarize our 
basic assumptions.
Firstly, we assume that there are three families of
leptons, which means that the charged-lepton mass matrix $M_\ell$, the
Dirac neutrino mass matrix $M_D$ and the Majorana neutrino mass matrix
$M_L$ are all $3 \times 3$ matrices.
Secondly, the tree-level mass
matrices which contain the texture zeros should be compatible with 
our knowledge about
the lepton masses and the
lepton mixing matrix, which leads us to the requirements 
$\text{rank}(M_\ell) = 3$ and 
$\text{rank}(M_D)\, \mbox{or}\ \text{rank}(M_L) \geq 2$
and excludes some textures in $M_L$ with degenerate neutrino masses.
Thirdly, 
the gauge-multiplet structure of the lepton fields is the same as in
the Standard Model, \textit{i.e.}\ left-handed doublets and right-handed
singlets. This last assumption fixes the allowed forms of weak-basis
transformations; for instance, in the case of Dirac neutrinos we are
allowed to perform independent weak-basis transformations on the
right-handed charged-lepton fields and the right-handed neutrino
fields.

One of the main points of this paper is that we admit all possible 
combinations of texture zeros in the pairs of mass matrices 
$(M_\ell,M_D)$ and $(M_\ell,M_L)$. In particular, we allow for
non-diagonal $M_\ell$. Since
there is a huge number of such combinations even after taking into
account the second assumption above, we had to address the problem of
equivalent texture zeros in the pairs of mass matrices, 
\textit{i.e.}\ textures which are related through
weak-basis transformations with permutation matrices (weak-basis
permutations). We have found 570 inequivalent classes of texture zeros
in the Dirac case and 298 classes in the Majorana case;
for both cases about 75\% of the classes are compatible with the data.
However,
if we consider only
classes which are maximally restrictive---\textit{cf.}
section~\ref{family-tree}, we are down to about 30 classes of texture
zeros for each of the four categories defined by 
Dirac/Majorana nature and normal/inverted 
ordering of the neutrino mass spectrum---see
table~\ref{chisquared-results}. 

We have also attempted to identify the predictive classes of texture zeros
by defining numerical measures of predictivity in
section~\ref{predictivity}. However, applying these measures to the
eight experimentally known observables, namely 
charged-lepton masses, neutrino mass-squared differences and mixing angles, 
it turned out that, for all but one of the models, none of these eight
observables can be predicted by using the values of the other seven
observables as input. On the other hand, using the values of all eight
known observables it is possible to predict in almost all viable and
maximally restrictive classes of texture zeros the smallest neutrino
mass $m_0$ and thus the absolute neutrino mass scale. However, even
this has a rather trivial explanation: most of these classes of
texture zeros predict $m_0 = 0$, but in these cases the neutrino mass matrix
has always rank two. The small rank of most neutrino mass matrices is
simply the consequence that we consider maximally restrictive classes. 
The main results of this paper are summarized in
tables~\ref{Ml-MD-normal} to~\ref{Ml-MnuL-inverted}.

In summary, pure texture zero models are astonishingly weak in their
predictions. This also holds for the hitherto neglected scenarios
where $M_\ell$ is non-diagonal. Of course, 
it could be that texture zero models which we have excluded
in this work because they fail at the tree level become compatible
with the data and are predictive when radiative corrections are taken
into account. Apart from this loop hole, we rather draw the
conclusion that predictive mass matrices
need also relations among the non-zero
matrix elements.

\paragraph{Acknowledgments:} This work is supported by the Austrian
Science Fund (FWF), Project No.\ P~24161-N16. P.O.L. thanks 
Helmut Moser for his tireless servicing of our research group's computer cluster
on which the
computations for this work were performed.

\begin{appendix}

\section{Details on the numerical analysis}\label{details-numerics}

All numerical methods used in this work are based on
minimization or maximization of functions of real parameters
$x \in \mathbbm{R}^n$. Since maximization of a real-valued
function $f(x)$ can be achieved by minimizing $-f(x)$,
we only need a minimization algorithm.
Our algorithm of choice is the Nelder-Mead downhill simplex
algorithm~\cite{nelder-mead}, which we implemented in
the programming language C~\cite{C-language}.\footnote{We also
wrote an independent program for the $\chi^2$-analysis
in MATLAB\textregistered~\cite{matlab}
using the built-in function \texttt{fminsearch}. We randomly
picked some types of texture zeros and compared the minima
of $\chi^2(x)$ found with our C-program and the MATLAB\textregistered-program.
Within the numerical accuracy, the found minima coincided.}
To increase the quality of the 
minima,
the algorithm was started 
repeatedly
with different
``start simplices''.\footnote{The Nelder-Mead algorithm
for function minimization in $n$ dimensions is
based on manipulation of the $n+1$ vertices of
a simplex in $\mathbbm{R}^n$.}
In order to further
improve the quality of the 
minima, 
once a
local minimum of $f(x)$ was found, 
we made a small perturbation
around the minimum and restarted the algorithm.
This algorithm ``Nelder-Mead + perturbations''
was implemented and used as described in appendix A of~\cite{PhD-Ludl}.
For the minimization of some functions, we limited the total computation
time. The details on the minimization procedures can
be found in table~\ref{table-details}.

Since the Nelder-Mead algorithm is itself not capable
of respecting constraints on the domain of the function $f$
to be minimized, all constraints must be directly 
incorporated
into
the definition of $f$. For example, in all our numerical
studies in this paper, we took into account the cosmological
constraint on the sum of the
neutrino masses~\cite{PDG}
\begin{equation}
	\sum_{k=1}^3 m_k < O(1\,\text{eV}),
\end{equation}
where the concrete number on the right-hand side of this equation
depends on the analysis and the data set.
For definiteness, we replaced each function $f(x)$ to be minimized by
\begin{equation}
\hat{f}(x) \equiv
 \begin{cases}
     10^9 & \text{if } \sum_{k=1}^3 m_k > 1\,\text{eV}, \\
     f(x) & \text{else},
 \end{cases}
\end{equation}
\textit{i.e.}\ we imposed the constraint of equation~(\ref{cosmo}).
The high value of $10^9$ 
enforces 
that the global minimum of $\hat{f}(x)$
respects that constraint.

\paragraph{Definitions of the functions relevant for our analysis:} The
$\chi^2$-function was implemented as defined in
equation~(\ref{def-chisquared}). The computation of
the predictivity measure $\Delta(\mathcal{O}_i)$ given
in equation~(\ref{predictivity-measure}) was done by
minimizing
\begin{equation}
\Delta_i(x) \equiv
\begin{cases}
 10^6 +\widetilde{\chi}_i^2(x) & \text{ if } \widetilde{\chi}_i^2(x) >
 \chi^2_\mathrm{min} + \delta\chi^2 \text{ or } \chi_j^2(x) > 25 \quad
 (j\neq i),
\\ 
 -\chi_i^2(x) & \text{else}.
\end{cases}
\end{equation}
The high value of $10^6$ ensures 
that 
the minimum of $\Delta_i(x)$ 
respects the
constraint $x\in B_i$ where $B_i$ is defined in equation~(\ref{def-B_i}). 
Adding $\chi_i^2(x)$ to
$10^6$ in the first line of the above equation drives the simplex
towards a region respecting the desired constraints.
The optimizations in equations~(\ref{optim-ml}) and~(\ref{optim-O})
were done using the same technique.

\renewcommand{\arraystretch}{1.4}
\begin{sidewaystable}
\begin{center}
\begin{tabular}{|l|c|c|c|l|}
\hline
Analysis & $\delta\chi^2$ & max.\ no.\ of random start simplices & stop computation if & time limit\\
\hline
Minimization of $\chi^2(x)$ & --- & $1000$ & $\chi^2(x)<10^{-4}$ & no time limit\\
Minimization of $\Delta_i(x)$ & 
$0.1$ & $10^6$ & $\Delta_i(x) < -100$ & 2 hours\\
Computation of range of $m_0$ & $0.1$ & $10^6$ & $m_0^\mathrm{min} < 10^{-3}\,\text{eV}$ & 2 hours\\
Computation of ranges of $\delta$, $\rho$, $\sigma$ & $0.1$ & $10^6$ & $\mathcal{O}^\mathrm{min} < 0.01$, $\mathcal{O}^\mathrm{max} > 2\pi -0.01$ & 2 hours\\
Computation of $m_\ell^\text{min}$ and $m_\ell^\text{max}$ & $0.1$ & $10^6$ & $m_\ell^\mathrm{min}< 0.01\,\text{MeV}$, $m_\ell^\mathrm{max}>10\,\text{GeV}$ & 2 hours\\
Computation of $m_{\beta\beta}^\text{min}$ and $m_{\beta\beta}^\text{max}$ & $0.1$ & $10^6$ & $m_{\beta\beta}^\mathrm{min} < 10^{-3}\,\text{eV}$ & 2 hours\\
\hline
\end{tabular}
\end{center}
\caption{Details on the numerical analysis carried out for this paper.
In line~2 the time limit is per texture and observable, in the other
lines the limit is per texture and minimization and the same for
maximization.}\label{table-details} 
\end{sidewaystable}
\renewcommand{\arraystretch}{1.0}

\end{appendix}

\end{document}

%% file: Mlclasses-table.tex
\begin{tabular}{|llll|}
\hline
$3_1^{(\ell)}\sim$
$\begin{pmatrix}
0 & 0 & \times \\
\times & \times & 0 \\
\times & \times & \times\end{pmatrix}$
&
$3_2^{(\ell)}\sim$
$\begin{pmatrix}
0 & \times & \times \\
0 & \times & \times \\
\times & 0 & \times\end{pmatrix}$
&
$3_3^{(\ell)}\sim$
$\begin{pmatrix}
0 & \times & \times \\
\times & 0 & \times \\
\times & \times & 0\end{pmatrix}$
&
\\
\hline
$4_1^{(\ell)}\sim$
$\begin{pmatrix}
0 & 0 & \times \\
0 & \times & 0 \\
\times & \times & \times\end{pmatrix}$
&
$4_2^{(\ell)}\sim$
$\begin{pmatrix}
0 & 0 & \times \\
0 & \times & \times \\
\times & 0 & \times\end{pmatrix}$
&
$4_3^{(\ell)}\sim$
$\begin{pmatrix}
0 & 0 & \times \\
0 & \times & \times \\
\times & \times & 0\end{pmatrix}$
&
$4_4^{(\ell)}\sim$
$\begin{pmatrix}
0 & 0 & \times \\
\times & \times & 0 \\
\times & \times & 0\end{pmatrix}$
\\
\hline
$5_1^{(\ell)}\sim$
$\begin{pmatrix}
0 & 0 & \times \\
0 & \times & 0 \\
\times & 0 & \times\end{pmatrix}$
&
&
&
\\
\hline
$6_1^{(\ell)}\sim$
$\begin{pmatrix}
0 & 0 & \times \\
0 & \times & 0 \\
\times & 0 & 0\end{pmatrix}$
&
&
&
\\
\hline
\end{tabular}

%% file: MDclasses-table1.tex
\begin{tabular}{|llll|}
\hline
$2_{1}^{(\nu_D)}\sim$
$\begin{pmatrix}
0 & \times & 0 \\
\times & \times & \times \\
\times & \times & \times\end{pmatrix}$
&
$2_{2}^{(\nu_D)}\sim$
$\begin{pmatrix}
0 & \times & \times \\
\times & \times & 0 \\
\times & \times & \times\end{pmatrix}$
&
$2_{3}^{(\nu_D)}\sim$
$\begin{pmatrix}
\times & 0 & \times \\
\times & 0 & \times \\
\times & \times & \times\end{pmatrix}$
&
\\
\hline
$3_{1}^{(\nu_D)}\sim$
$\begin{pmatrix}
0 & 0 & 0 \\
\times & \times & \times \\
\times & \times & \times\end{pmatrix}$
&
$3_{2}^{(\nu_D)}\sim$
$\begin{pmatrix}
0 & 0 & \times \\
\times & 0 & \times \\
\times & \times & \times\end{pmatrix}$
&
$3_{3}^{(\nu_D)}\sim$
$\begin{pmatrix}
0 & 0 & \times \\
\times & \times & 0 \\
\times & \times & \times\end{pmatrix}$
&
$3_{4}^{(\nu_D)}\sim$
$\begin{pmatrix}
0 & \times & 0 \\
0 & \times & \times \\
\times & \times & \times\end{pmatrix}$
\\
$3_{5}^{(\nu_D)}\sim$
$\begin{pmatrix}
0 & \times & 0 \\
\times & 0 & \times \\
\times & \times & \times\end{pmatrix}$
&
$3_{6}^{(\nu_D)}\sim$
$\begin{pmatrix}
0 & \times & 0 \\
\times & \times & 0 \\
\times & \times & \times\end{pmatrix}$
&
$3_{7}^{(\nu_D)}\sim$
$\begin{pmatrix}
0 & \times & \times \\
0 & \times & \times \\
0 & \times & \times\end{pmatrix}$
&
$3_{8}^{(\nu_D)}\sim$
$\begin{pmatrix}
0 & \times & \times \\
0 & \times & \times \\
\times & 0 & \times\end{pmatrix}$
\\
$3_{9}^{(\nu_D)}\sim$
$\begin{pmatrix}
0 & \times & \times \\
0 & \times & \times \\
\times & \times & 0\end{pmatrix}$
&
$3_{10}^{(\nu_D)}\sim$
$\begin{pmatrix}
0 & \times & \times \\
\times & 0 & 0 \\
\times & \times & \times\end{pmatrix}$
&
$3_{11}^{(\nu_D)}\sim$
$\begin{pmatrix}
0 & \times & \times \\
\times & 0 & \times \\
\times & 0 & \times\end{pmatrix}$
&
$3_{12}^{(\nu_D)}\sim$
$\begin{pmatrix}
0 & \times & \times \\
\times & 0 & \times \\
\times & \times & 0\end{pmatrix}$
\\
$3_{13}^{(\nu_D)}\sim$
$\begin{pmatrix}
0 & \times & \times \\
\times & \times & 0 \\
\times & \times & 0\end{pmatrix}$
&
$3_{14}^{(\nu_D)}\sim$
$\begin{pmatrix}
\times & 0 & 0 \\
\times & 0 & \times \\
\times & \times & \times\end{pmatrix}$
&
$3_{15}^{(\nu_D)}\sim$
$\begin{pmatrix}
\times & 0 & \times \\
\times & 0 & \times \\
\times & 0 & \times\end{pmatrix}$
&
$3_{16}^{(\nu_D)}\sim$
$\begin{pmatrix}
\times & 0 & \times \\
\times & 0 & \times \\
\times & \times & 0\end{pmatrix}$
\\
$3_{17}^{(\nu_D)}\sim$
$\begin{pmatrix}
\times & 0 & \times \\
\times & \times & 0 \\
\times & \times & 0\end{pmatrix}$
&
$3_{18}^{(\nu_D)}\sim$
$\begin{pmatrix}
\times & \times & 0 \\
\times & \times & 0 \\
\times & \times & 0\end{pmatrix}$
&
&
\\
\hline
$4_{1}^{(\nu_D)}\sim$
$\begin{pmatrix}
0 & 0 & 0 \\
0 & \times & \times \\
\times & \times & \times\end{pmatrix}$
&
$4_{2}^{(\nu_D)}\sim$
$\begin{pmatrix}
0 & 0 & 0 \\
\times & 0 & \times \\
\times & \times & \times\end{pmatrix}$
&
$4_{3}^{(\nu_D)}\sim$
$\begin{pmatrix}
0 & 0 & 0 \\
\times & \times & 0 \\
\times & \times & \times\end{pmatrix}$
&
$4_{4}^{(\nu_D)}\sim$
$\begin{pmatrix}
0 & 0 & \times \\
0 & 0 & \times \\
\times & \times & \times\end{pmatrix}$
\\
$4_{5}^{(\nu_D)}\sim$
$\begin{pmatrix}
0 & 0 & \times \\
0 & \times & 0 \\
\times & \times & \times\end{pmatrix}$
&
$4_{6}^{(\nu_D)}\sim$
$\begin{pmatrix}
0 & 0 & \times \\
0 & \times & \times \\
0 & \times & \times\end{pmatrix}$
&
$4_{7}^{(\nu_D)}\sim$
$\begin{pmatrix}
0 & 0 & \times \\
0 & \times & \times \\
\times & 0 & \times\end{pmatrix}$
&
$4_{8}^{(\nu_D)}\sim$
$\begin{pmatrix}
0 & 0 & \times \\
0 & \times & \times \\
\times & \times & 0\end{pmatrix}$
\\
$4_{9}^{(\nu_D)}\sim$
$\begin{pmatrix}
0 & 0 & \times \\
\times & 0 & 0 \\
\times & \times & \times\end{pmatrix}$
&
$4_{10}^{(\nu_D)}\sim$
$\begin{pmatrix}
0 & 0 & \times \\
\times & 0 & \times \\
\times & 0 & \times\end{pmatrix}$
&
$4_{11}^{(\nu_D)}\sim$
$\begin{pmatrix}
0 & 0 & \times \\
\times & 0 & \times \\
\times & \times & 0\end{pmatrix}$
&
$4_{12}^{(\nu_D)}\sim$
$\begin{pmatrix}
0 & 0 & \times \\
\times & \times & 0 \\
\times & \times & 0\end{pmatrix}$
\\
$4_{13}^{(\nu_D)}\sim$
$\begin{pmatrix}
0 & \times & 0 \\
0 & \times & 0 \\
\times & \times & \times\end{pmatrix}$
&
$4_{14}^{(\nu_D)}\sim$
$\begin{pmatrix}
0 & \times & 0 \\
0 & \times & \times \\
0 & \times & \times\end{pmatrix}$
&
$4_{15}^{(\nu_D)}\sim$
$\begin{pmatrix}
0 & \times & 0 \\
0 & \times & \times \\
\times & 0 & \times\end{pmatrix}$
&
$4_{16}^{(\nu_D)}\sim$
$\begin{pmatrix}
0 & \times & 0 \\
0 & \times & \times \\
\times & \times & 0\end{pmatrix}$
\\
$4_{17}^{(\nu_D)}\sim$
$\begin{pmatrix}
0 & \times & 0 \\
\times & 0 & 0 \\
\times & \times & \times\end{pmatrix}$
&
$4_{18}^{(\nu_D)}\sim$
$\begin{pmatrix}
0 & \times & 0 \\
\times & 0 & \times \\
\times & 0 & \times\end{pmatrix}$
&
$4_{19}^{(\nu_D)}\sim$
$\begin{pmatrix}
0 & \times & 0 \\
\times & 0 & \times \\
\times & \times & 0\end{pmatrix}$
&
$4_{20}^{(\nu_D)}\sim$
$\begin{pmatrix}
0 & \times & 0 \\
\times & \times & 0 \\
\times & \times & 0\end{pmatrix}$
\\
$4_{21}^{(\nu_D)}\sim$
$\begin{pmatrix}
0 & \times & \times \\
0 & \times & \times \\
\times & 0 & 0\end{pmatrix}$
&
$4_{22}^{(\nu_D)}\sim$
$\begin{pmatrix}
0 & \times & \times \\
\times & 0 & 0 \\
\times & 0 & \times\end{pmatrix}$
&
$4_{23}^{(\nu_D)}\sim$
$\begin{pmatrix}
0 & \times & \times \\
\times & 0 & 0 \\
\times & \times & 0\end{pmatrix}$
&
$4_{24}^{(\nu_D)}\sim$
$\begin{pmatrix}
\times & 0 & 0 \\
\times & 0 & 0 \\
\times & \times & \times\end{pmatrix}$
\\
$4_{25}^{(\nu_D)}\sim$
$\begin{pmatrix}
\times & 0 & 0 \\
\times & 0 & \times \\
\times & 0 & \times\end{pmatrix}$
&
$4_{26}^{(\nu_D)}\sim$
$\begin{pmatrix}
\times & 0 & 0 \\
\times & 0 & \times \\
\times & \times & 0\end{pmatrix}$
&
$4_{27}^{(\nu_D)}\sim$
$\begin{pmatrix}
\times & 0 & 0 \\
\times & \times & 0 \\
\times & \times & 0\end{pmatrix}$
&
\\
\hline
\end{tabular}

%% file: MDclasses-table2.tex
\begin{tabular}{|llll|}
\hline
$5_{1}^{(\nu_D)}\sim$
$\begin{pmatrix}
0 & 0 & 0 \\
0 & 0 & \times \\
\times & \times & \times\end{pmatrix}$
&
$5_{2}^{(\nu_D)}\sim$
$\begin{pmatrix}
0 & 0 & 0 \\
0 & \times & 0 \\
\times & \times & \times\end{pmatrix}$
&
$5_{3}^{(\nu_D)}\sim$
$\begin{pmatrix}
0 & 0 & 0 \\
0 & \times & \times \\
0 & \times & \times\end{pmatrix}$
&
$5_{4}^{(\nu_D)}\sim$
$\begin{pmatrix}
0 & 0 & 0 \\
0 & \times & \times \\
\times & 0 & \times\end{pmatrix}$
\\
$5_{5}^{(\nu_D)}\sim$
$\begin{pmatrix}
0 & 0 & 0 \\
0 & \times & \times \\
\times & \times & 0\end{pmatrix}$
&
$5_{6}^{(\nu_D)}\sim$
$\begin{pmatrix}
0 & 0 & 0 \\
\times & 0 & 0 \\
\times & \times & \times\end{pmatrix}$
&
$5_{7}^{(\nu_D)}\sim$
$\begin{pmatrix}
0 & 0 & 0 \\
\times & 0 & \times \\
\times & 0 & \times\end{pmatrix}$
&
$5_{8}^{(\nu_D)}\sim$
$\begin{pmatrix}
0 & 0 & 0 \\
\times & 0 & \times \\
\times & \times & 0\end{pmatrix}$
\\
$5_{9}^{(\nu_D)}\sim$
$\begin{pmatrix}
0 & 0 & 0 \\
\times & \times & 0 \\
\times & \times & 0\end{pmatrix}$
&
$5_{10}^{(\nu_D)}\sim$
$\begin{pmatrix}
0 & 0 & \times \\
0 & 0 & \times \\
0 & \times & \times\end{pmatrix}$
&
$5_{11}^{(\nu_D)}\sim$
$\begin{pmatrix}
0 & 0 & \times \\
0 & 0 & \times \\
\times & 0 & \times\end{pmatrix}$
&
$5_{12}^{(\nu_D)}\sim$
$\begin{pmatrix}
0 & 0 & \times \\
0 & 0 & \times \\
\times & \times & 0\end{pmatrix}$
\\
$5_{13}^{(\nu_D)}\sim$
$\begin{pmatrix}
0 & 0 & \times \\
0 & \times & 0 \\
0 & \times & \times\end{pmatrix}$
&
$5_{14}^{(\nu_D)}\sim$
$\begin{pmatrix}
0 & 0 & \times \\
0 & \times & 0 \\
\times & 0 & \times\end{pmatrix}$
&
$5_{15}^{(\nu_D)}\sim$
$\begin{pmatrix}
0 & 0 & \times \\
0 & \times & 0 \\
\times & \times & 0\end{pmatrix}$
&
$5_{16}^{(\nu_D)}\sim$
$\begin{pmatrix}
0 & 0 & \times \\
0 & \times & \times \\
\times & 0 & 0\end{pmatrix}$
\\
$5_{17}^{(\nu_D)}\sim$
$\begin{pmatrix}
0 & 0 & \times \\
\times & 0 & 0 \\
\times & 0 & \times\end{pmatrix}$
&
$5_{18}^{(\nu_D)}\sim$
$\begin{pmatrix}
0 & 0 & \times \\
\times & 0 & 0 \\
\times & \times & 0\end{pmatrix}$
&
$5_{19}^{(\nu_D)}\sim$
$\begin{pmatrix}
0 & \times & 0 \\
0 & \times & 0 \\
0 & \times & \times\end{pmatrix}$
&
$5_{20}^{(\nu_D)}\sim$
$\begin{pmatrix}
0 & \times & 0 \\
0 & \times & 0 \\
\times & 0 & \times\end{pmatrix}$
\\
$5_{21}^{(\nu_D)}\sim$
$\begin{pmatrix}
0 & \times & 0 \\
0 & \times & 0 \\
\times & \times & 0\end{pmatrix}$
&
$5_{22}^{(\nu_D)}\sim$
$\begin{pmatrix}
0 & \times & 0 \\
0 & \times & \times \\
\times & 0 & 0\end{pmatrix}$
&
$5_{23}^{(\nu_D)}\sim$
$\begin{pmatrix}
0 & \times & 0 \\
\times & 0 & 0 \\
\times & 0 & \times\end{pmatrix}$
&
$5_{24}^{(\nu_D)}\sim$
$\begin{pmatrix}
0 & \times & 0 \\
\times & 0 & 0 \\
\times & \times & 0\end{pmatrix}$
\\
$5_{25}^{(\nu_D)}\sim$
$\begin{pmatrix}
0 & \times & \times \\
\times & 0 & 0 \\
\times & 0 & 0\end{pmatrix}$
&
$5_{26}^{(\nu_D)}\sim$
$\begin{pmatrix}
\times & 0 & 0 \\
\times & 0 & 0 \\
\times & 0 & \times\end{pmatrix}$
&
$5_{27}^{(\nu_D)}\sim$
$\begin{pmatrix}
\times & 0 & 0 \\
\times & 0 & 0 \\
\times & \times & 0\end{pmatrix}$
&
\\
\hline
$6_{1}^{(\nu_D)}\sim$
$\begin{pmatrix}
0 & 0 & 0 \\
0 & 0 & \times \\
0 & \times & \times\end{pmatrix}$
&
$6_{2}^{(\nu_D)}\sim$
$\begin{pmatrix}
0 & 0 & 0 \\
0 & 0 & \times \\
\times & 0 & \times\end{pmatrix}$
&
$6_{3}^{(\nu_D)}\sim$
$\begin{pmatrix}
0 & 0 & 0 \\
0 & 0 & \times \\
\times & \times & 0\end{pmatrix}$
&
$6_{4}^{(\nu_D)}\sim$
$\begin{pmatrix}
0 & 0 & 0 \\
0 & \times & 0 \\
0 & \times & \times\end{pmatrix}$
\\
$6_{5}^{(\nu_D)}\sim$
$\begin{pmatrix}
0 & 0 & 0 \\
0 & \times & 0 \\
\times & 0 & \times\end{pmatrix}$
&
$6_{6}^{(\nu_D)}\sim$
$\begin{pmatrix}
0 & 0 & 0 \\
0 & \times & 0 \\
\times & \times & 0\end{pmatrix}$
&
$6_{7}^{(\nu_D)}\sim$
$\begin{pmatrix}
0 & 0 & 0 \\
0 & \times & \times \\
\times & 0 & 0\end{pmatrix}$
&
$6_{8}^{(\nu_D)}\sim$
$\begin{pmatrix}
0 & 0 & 0 \\
\times & 0 & 0 \\
\times & 0 & \times\end{pmatrix}$
\\
$6_{9}^{(\nu_D)}\sim$
$\begin{pmatrix}
0 & 0 & 0 \\
\times & 0 & 0 \\
\times & \times & 0\end{pmatrix}$
&
$6_{10}^{(\nu_D)}\sim$
$\begin{pmatrix}
0 & 0 & \times \\
0 & 0 & \times \\
0 & \times & 0\end{pmatrix}$
&
$6_{11}^{(\nu_D)}\sim$
$\begin{pmatrix}
0 & 0 & \times \\
0 & 0 & \times \\
\times & 0 & 0\end{pmatrix}$
&
$6_{12}^{(\nu_D)}\sim$
$\begin{pmatrix}
0 & 0 & \times \\
0 & \times & 0 \\
0 & \times & 0\end{pmatrix}$
\\
$6_{13}^{(\nu_D)}\sim$
$\begin{pmatrix}
0 & 0 & \times \\
0 & \times & 0 \\
\times & 0 & 0\end{pmatrix}$
&
$6_{14}^{(\nu_D)}\sim$
$\begin{pmatrix}
0 & 0 & \times \\
\times & 0 & 0 \\
\times & 0 & 0\end{pmatrix}$
&
$6_{15}^{(\nu_D)}\sim$
$\begin{pmatrix}
0 & \times & 0 \\
0 & \times & 0 \\
\times & 0 & 0\end{pmatrix}$
&
$6_{16}^{(\nu_D)}\sim$
$\begin{pmatrix}
0 & \times & 0 \\
\times & 0 & 0 \\
\times & 0 & 0\end{pmatrix}$
\\
\hline
$7_{1}^{(\nu_D)}\sim$
$\begin{pmatrix}
0 & 0 & 0 \\
0 & 0 & \times \\
0 & \times & 0\end{pmatrix}$
&
$7_{2}^{(\nu_D)}\sim$
$\begin{pmatrix}
0 & 0 & 0 \\
0 & 0 & \times \\
\times & 0 & 0\end{pmatrix}$
&
$7_{3}^{(\nu_D)}\sim$
$\begin{pmatrix}
0 & 0 & 0 \\
0 & \times & 0 \\
\times & 0 & 0\end{pmatrix}$
&
\\
\hline
\end{tabular}

%% file: Ml-MD-classes_redundant.tex
\begin{tabular}{|c|lllllllllllllll|}
\hline
$M_\ell$ & \multicolumn{15}{|c|}{$M_D$} \\
\hline
$3_2$ & $3_{10}$ & $3_{11}$ & $3_{14}$ & $3_{15}$ & $3_{16}$ & $3_{17}$ & $4_{2}$ & $4_{9}$ & $4_{10}$ & $4_{11}$ & $4_{21}$ & $4_{22}$ & $4_{23}$ & $4_{24}$ & $4_{25}$ \\
 & $4_{26}$ & $4_{27}$ & $5_{6}$ & $5_{7}$ & $5_{8}$ & $5_{11}$ & $5_{16}$ & $5_{17}$ & $5_{18}$ & $5_{23}$ & $5_{25}$ & $5_{26}$ & $5_{27}$ & $6_{2}$ & $6_{7}$ \\
 & $6_{8}$ & $6_{9}$ & $6_{11}$ & $6_{14}$ & $6_{16}$ & $7_{2}$ &  &  &  &  &  &  &  &  & \\
\hline
$3_3$ & $3_{4}$ & $3_{5}$ & $3_{6}$ & $3_{9}$ & $3_{10}$ & $3_{11}$ & $3_{13}$ & $3_{14}$ & $3_{15}$ & $3_{16}$ & $3_{17}$ & $3_{18}$ & $4_{2}$ & $4_{3}$ & $4_{9}$ \\
 & $4_{10}$ & $4_{11}$ & $4_{13}$ & $4_{14}$ & $4_{15}$ & $4_{16}$ & $4_{17}$ & $4_{18}$ & $4_{19}$ & $4_{20}$ & $4_{21}$ & $4_{22}$ & $4_{23}$ & $4_{24}$ & $4_{25}$ \\
 & $4_{26}$ & $4_{27}$ & $5_{2}$ & $5_{5}$ & $5_{6}$ & $5_{7}$ & $5_{8}$ & $5_{9}$ & $5_{11}$ & $5_{15}$ & $5_{16}$ & $5_{17}$ & $5_{18}$ & $5_{19}$ & $5_{20}$ \\
 & $5_{21}$ & $5_{22}$ & $5_{23}$ & $5_{24}$ & $5_{25}$ & $5_{26}$ & $5_{27}$ & $6_{2}$ & $6_{4}$ & $6_{5}$ & $6_{6}$ & $6_{7}$ & $6_{8}$ & $6_{9}$ & $6_{11}$ \\
 & $6_{12}$ & $6_{14}$ & $6_{15}$ & $6_{16}$ & $7_{2}$ & $7_{3}$ &  &  &  &  &  &  &  &  & \\
\hline
$4_1$ & $3_{10}$ & $3_{11}$ & $3_{14}$ & $3_{15}$ & $3_{16}$ & $3_{17}$ & $4_{2}$ & $4_{9}$ & $4_{10}$ & $4_{11}$ & $4_{21}$ & $4_{22}$ & $4_{23}$ & $4_{24}$ & $4_{25}$ \\
 & $4_{26}$ & $4_{27}$ & $5_{6}$ & $5_{7}$ & $5_{8}$ & $5_{11}$ & $5_{16}$ & $5_{17}$ & $5_{18}$ & $5_{23}$ & $5_{25}$ & $5_{26}$ & $5_{27}$ & $6_{2}$ & $6_{7}$ \\
 & $6_{8}$ & $6_{9}$ & $6_{11}$ & $6_{14}$ & $6_{16}$ & $7_{2}$ &  &  &  &  &  &  &  &  & \\
\hline
$4_2$ & $3_{5}$ & $3_{6}$ & $3_{9}$ & $3_{13}$ & $3_{17}$ & $3_{18}$ & $4_{3}$ & $4_{13}$ & $4_{14}$ & $4_{15}$ & $4_{16}$ & $4_{17}$ & $4_{18}$ & $4_{19}$ & $4_{20}$ \\
 & $4_{23}$ & $4_{27}$ & $5_{2}$ & $5_{5}$ & $5_{9}$ & $5_{15}$ & $5_{19}$ & $5_{20}$ & $5_{21}$ & $5_{22}$ & $5_{23}$ & $5_{24}$ & $5_{27}$ & $6_{4}$ & $6_{5}$ \\
 & $6_{6}$ & $6_{9}$ & $6_{12}$ & $6_{15}$ & $6_{16}$ & $7_{3}$ &  &  &  &  &  &  &  &  & \\
\hline
$4_4$ & $3_{5}$ & $3_{6}$ & $3_{9}$ & $3_{13}$ & $3_{17}$ & $3_{18}$ & $4_{3}$ & $4_{13}$ & $4_{14}$ & $4_{15}$ & $4_{16}$ & $4_{17}$ & $4_{18}$ & $4_{19}$ & $4_{20}$ \\
 & $4_{23}$ & $4_{27}$ & $5_{2}$ & $5_{5}$ & $5_{9}$ & $5_{15}$ & $5_{19}$ & $5_{20}$ & $5_{21}$ & $5_{22}$ & $5_{23}$ & $5_{24}$ & $5_{27}$ & $6_{4}$ & $6_{5}$ \\
 & $6_{6}$ & $6_{9}$ & $6_{12}$ & $6_{15}$ & $6_{16}$ & $7_{3}$ &  &  &  &  &  &  &  &  & \\
\hline
$6_1$ & $3_{4}$ & $3_{5}$ & $3_{6}$ & $3_{9}$ & $3_{10}$ & $3_{11}$ & $3_{13}$ & $3_{14}$ & $3_{15}$ & $3_{16}$ & $3_{17}$ & $3_{18}$ & $4_{2}$ & $4_{3}$ & $4_{9}$ \\
 & $4_{10}$ & $4_{11}$ & $4_{13}$ & $4_{14}$ & $4_{15}$ & $4_{16}$ & $4_{17}$ & $4_{18}$ & $4_{19}$ & $4_{20}$ & $4_{21}$ & $4_{22}$ & $4_{23}$ & $4_{24}$ & $4_{25}$ \\
 & $4_{26}$ & $4_{27}$ & $5_{2}$ & $5_{5}$ & $5_{6}$ & $5_{7}$ & $5_{8}$ & $5_{9}$ & $5_{11}$ & $5_{15}$ & $5_{16}$ & $5_{17}$ & $5_{18}$ & $5_{19}$ & $5_{20}$ \\
 & $5_{21}$ & $5_{22}$ & $5_{23}$ & $5_{24}$ & $5_{25}$ & $5_{26}$ & $5_{27}$ & $6_{2}$ & $6_{4}$ & $6_{5}$ & $6_{6}$ & $6_{7}$ & $6_{8}$ & $6_{9}$ & $6_{11}$ \\
 & $6_{12}$ & $6_{14}$ & $6_{15}$ & $6_{16}$ & $7_{2}$ & $7_{3}$ &  &  &  &  &  &  &  &  & \\
\hline
\end{tabular}

%% file: MnuLclasses-table.tex
\begin{tabular}{|llll|}
\hline
$1_{1}^{(\nu_L)}\sim$
$\begin{pmatrix}
0 & \times & \times \\
\times & \times & \times \\
\times & \times & \times\end{pmatrix}$
&
$1_{2}^{(\nu_L)}\sim$
$\begin{pmatrix}
\times & \times & \times \\
\times & 0 & \times \\
\times & \times & \times\end{pmatrix}$
&
$1_{3}^{(\nu_L)}\sim$
$\begin{pmatrix}
\times & \times & \times \\
\times & \times & \times \\
\times & \times & 0\end{pmatrix}$
&
$1_{4}^{(\nu_L)}\sim$
$\begin{pmatrix}
\times & 0 & \times \\
0 & \times & \times \\
\times & \times & \times\end{pmatrix}$
\\
$1_{5}^{(\nu_L)}\sim$
$\begin{pmatrix}
\times & \times & 0 \\
\times & \times & \times \\
0 & \times & \times\end{pmatrix}$
&
$1_{6}^{(\nu_L)}\sim$
$\begin{pmatrix}
\times & \times & \times \\
\times & \times & 0 \\
\times & 0 & \times\end{pmatrix}$
&
&
\\
\hline
$2_{1}^{(\nu_L)}\sim$
$\begin{pmatrix}
0 & \times & \times \\
\times & 0 & \times \\
\times & \times & \times\end{pmatrix}$
&
$2_{2}^{(\nu_L)}\sim$
$\begin{pmatrix}
0 & \times & \times \\
\times & \times & \times \\
\times & \times & 0\end{pmatrix}$
&
$2_{3}^{(\nu_L)}\sim$
$\begin{pmatrix}
\times & \times & \times \\
\times & 0 & \times \\
\times & \times & 0\end{pmatrix}$
&
$2_{4}^{(\nu_L)}\sim$
$\begin{pmatrix}
0 & 0 & \times \\
0 & \times & \times \\
\times & \times & \times\end{pmatrix}$
\\
$2_{5}^{(\nu_L)}\sim$
$\begin{pmatrix}
0 & \times & 0 \\
\times & \times & \times \\
0 & \times & \times\end{pmatrix}$
&
$2_{6}^{(\nu_L)}\sim$
$\begin{pmatrix}
0 & \times & \times \\
\times & \times & 0 \\
\times & 0 & \times\end{pmatrix}$
&
$2_{7}^{(\nu_L)}\sim$
$\begin{pmatrix}
\times & 0 & \times \\
0 & 0 & \times \\
\times & \times & \times\end{pmatrix}$
&
$2_{8}^{(\nu_L)}\sim$
$\begin{pmatrix}
\times & 0 & \times \\
0 & \times & \times \\
\times & \times & 0\end{pmatrix}$
\\
$2_{9}^{(\nu_L)}\sim$
$\begin{pmatrix}
\times & \times & 0 \\
\times & 0 & \times \\
0 & \times & \times\end{pmatrix}$
&
$2_{10}^{(\nu_L)}\sim$
$\begin{pmatrix}
\times & \times & 0 \\
\times & \times & \times \\
0 & \times & 0\end{pmatrix}$
&
$2_{11}^{(\nu_L)}\sim$
$\begin{pmatrix}
\times & \times & \times \\
\times & 0 & 0 \\
\times & 0 & \times\end{pmatrix}$
&
$2_{12}^{(\nu_L)}\sim$
$\begin{pmatrix}
\times & \times & \times \\
\times & \times & 0 \\
\times & 0 & 0\end{pmatrix}$
\\
$2_{13}^{(\nu_L)}\sim$
$\begin{pmatrix}
\times & 0 & 0 \\
0 & \times & \times \\
0 & \times & \times\end{pmatrix}$
&
$2_{14}^{(\nu_L)}\sim$
$\begin{pmatrix}
\times & 0 & \times \\
0 & \times & 0 \\
\times & 0 & \times\end{pmatrix}$
&
$2_{15}^{(\nu_L)}\sim$
$\begin{pmatrix}
\times & \times & 0 \\
\times & \times & 0 \\
0 & 0 & \times\end{pmatrix}$
&
\\
\hline
$3_{1}^{(\nu_L)}\sim$
$\begin{pmatrix}
0 & \times & \times \\
\times & 0 & \times \\
\times & \times & 0\end{pmatrix}$
&
$3_{2}^{(\nu_L)}\sim$
$\begin{pmatrix}
0 & 0 & \times \\
0 & 0 & \times \\
\times & \times & \times\end{pmatrix}$
&
$3_{3}^{(\nu_L)}\sim$
$\begin{pmatrix}
0 & 0 & \times \\
0 & \times & \times \\
\times & \times & 0\end{pmatrix}$
&
$3_{4}^{(\nu_L)}\sim$
$\begin{pmatrix}
0 & \times & 0 \\
\times & 0 & \times \\
0 & \times & \times\end{pmatrix}$
\\
$3_{5}^{(\nu_L)}\sim$
$\begin{pmatrix}
0 & \times & 0 \\
\times & \times & \times \\
0 & \times & 0\end{pmatrix}$
&
$3_{6}^{(\nu_L)}\sim$
$\begin{pmatrix}
0 & \times & \times \\
\times & 0 & 0 \\
\times & 0 & \times\end{pmatrix}$
&
$3_{7}^{(\nu_L)}\sim$
$\begin{pmatrix}
0 & \times & \times \\
\times & \times & 0 \\
\times & 0 & 0\end{pmatrix}$
&
$3_{8}^{(\nu_L)}\sim$
$\begin{pmatrix}
\times & 0 & \times \\
0 & 0 & \times \\
\times & \times & 0\end{pmatrix}$
\\
$3_{9}^{(\nu_L)}\sim$
$\begin{pmatrix}
\times & \times & 0 \\
\times & 0 & \times \\
0 & \times & 0\end{pmatrix}$
&
$3_{10}^{(\nu_L)}\sim$
$\begin{pmatrix}
\times & \times & \times \\
\times & 0 & 0 \\
\times & 0 & 0\end{pmatrix}$
&
$3_{11}^{(\nu_L)}\sim$
$\begin{pmatrix}
0 & 0 & 0 \\
0 & \times & \times \\
0 & \times & \times\end{pmatrix}$
&
$3_{12}^{(\nu_L)}\sim$
$\begin{pmatrix}
0 & 0 & \times \\
0 & \times & 0 \\
\times & 0 & \times\end{pmatrix}$
\\
$3_{13}^{(\nu_L)}\sim$
$\begin{pmatrix}
0 & \times & 0 \\
\times & \times & 0 \\
0 & 0 & \times\end{pmatrix}$
&
$3_{14}^{(\nu_L)}\sim$
$\begin{pmatrix}
\times & 0 & 0 \\
0 & 0 & \times \\
0 & \times & \times\end{pmatrix}$
&
$3_{15}^{(\nu_L)}\sim$
$\begin{pmatrix}
\times & 0 & 0 \\
0 & \times & \times \\
0 & \times & 0\end{pmatrix}$
&
$3_{16}^{(\nu_L)}\sim$
$\begin{pmatrix}
\times & 0 & \times \\
0 & 0 & 0 \\
\times & 0 & \times\end{pmatrix}$
\\
$3_{17}^{(\nu_L)}\sim$
$\begin{pmatrix}
\times & 0 & \times \\
0 & \times & 0 \\
\times & 0 & 0\end{pmatrix}$
&
$3_{18}^{(\nu_L)}\sim$
$\begin{pmatrix}
\times & \times & 0 \\
\times & 0 & 0 \\
0 & 0 & \times\end{pmatrix}$
&
$3_{19}^{(\nu_L)}\sim$
$\begin{pmatrix}
\times & \times & 0 \\
\times & \times & 0 \\
0 & 0 & 0\end{pmatrix}$
&
$3_{20}^{(\nu_L)}\sim$
$\begin{pmatrix}
\times & 0 & 0 \\
0 & \times & 0 \\
0 & 0 & \times\end{pmatrix}$
\\
\hline
$4_{1}^{(\nu_L)}\sim$
$\begin{pmatrix}
0 & 0 & 0 \\
0 & 0 & \times \\
0 & \times & \times\end{pmatrix}$
&
$4_{2}^{(\nu_L)}\sim$
$\begin{pmatrix}
0 & 0 & 0 \\
0 & \times & \times \\
0 & \times & 0\end{pmatrix}$
&
$4_{3}^{(\nu_L)}\sim$
$\begin{pmatrix}
0 & 0 & \times \\
0 & 0 & 0 \\
\times & 0 & \times\end{pmatrix}$
&
$4_{4}^{(\nu_L)}\sim$
$\begin{pmatrix}
0 & \times & 0 \\
\times & \times & 0 \\
0 & 0 & 0\end{pmatrix}$
\\
$4_{5}^{(\nu_L)}\sim$
$\begin{pmatrix}
\times & 0 & \times \\
0 & 0 & 0 \\
\times & 0 & 0\end{pmatrix}$
&
$4_{6}^{(\nu_L)}\sim$
$\begin{pmatrix}
\times & \times & 0 \\
\times & 0 & 0 \\
0 & 0 & 0\end{pmatrix}$
&
$4_{7}^{(\nu_L)}\sim$
$\begin{pmatrix}
0 & 0 & 0 \\
0 & \times & 0 \\
0 & 0 & \times\end{pmatrix}$
&
$4_{8}^{(\nu_L)}\sim$
$\begin{pmatrix}
\times & 0 & 0 \\
0 & 0 & 0 \\
0 & 0 & \times\end{pmatrix}$
\\
$4_{9}^{(\nu_L)}\sim$
$\begin{pmatrix}
\times & 0 & 0 \\
0 & \times & 0 \\
0 & 0 & 0\end{pmatrix}$
&
&
&
\\
\hline
\end{tabular}

%% file: Ml-MnuL-classes_redundant.tex
\begin{tabular}{|c|lllllllllllllll|}
\hline
$M_\ell$ & \multicolumn{15}{|c|}{$M_L$} \\
\hline
$3_2$ & $1_{2}$ & $1_{6}$ & $2_{3}$ & $2_{7}$ & $2_{9}$ & $2_{11}$ & $2_{12}$ & $2_{14}$ & $3_{6}$ & $3_{8}$ & $3_{9}$ & $3_{10}$ & $3_{14}$ & $3_{16}$ & $3_{17}$ \\
 & $3_{18}$ & $4_{3}$ & $4_{5}$ & $4_{6}$ & $4_{8}$ &  &  &  &  &  &  &  &  &  & \\
\hline
$3_3$ & $1_{2}$ & $1_{3}$ & $1_{5}$ & $1_{6}$ & $2_{2}$ & $2_{3}$ & $2_{5}$ & $2_{7}$ & $2_{8}$ & $2_{9}$ & $2_{10}$ & $2_{11}$ & $2_{12}$ & $2_{14}$ & $2_{15}$ \\
 & $3_{4}$ & $3_{5}$ & $3_{6}$ & $3_{7}$ & $3_{8}$ & $3_{9}$ & $3_{10}$ & $3_{13}$ & $3_{14}$ & $3_{15}$ & $3_{16}$ & $3_{17}$ & $3_{18}$ & $3_{19}$ & $4_{2}$ \\
 & $4_{3}$ & $4_{4}$ & $4_{5}$ & $4_{6}$ & $4_{8}$ & $4_{9}$ &  &  &  &  &  &  &  &  & \\
\hline
$4_1$ & $1_{2}$ & $1_{6}$ & $2_{3}$ & $2_{7}$ & $2_{9}$ & $2_{11}$ & $2_{12}$ & $2_{14}$ & $3_{6}$ & $3_{8}$ & $3_{9}$ & $3_{10}$ & $3_{14}$ & $3_{16}$ & $3_{17}$ \\
 & $3_{18}$ & $4_{3}$ & $4_{5}$ & $4_{6}$ & $4_{8}$ &  &  &  &  &  &  &  &  &  & \\
\hline
$4_2$ & $1_{3}$ & $1_{5}$ & $2_{2}$ & $2_{5}$ & $2_{9}$ & $2_{10}$ & $2_{12}$ & $2_{15}$ & $3_{4}$ & $3_{5}$ & $3_{7}$ & $3_{9}$ & $3_{13}$ & $3_{15}$ & $3_{18}$ \\
 & $3_{19}$ & $4_{2}$ & $4_{4}$ & $4_{6}$ & $4_{9}$ &  &  &  &  &  &  &  &  &  & \\
\hline
$4_4$ & $1_{3}$ & $1_{5}$ & $2_{2}$ & $2_{5}$ & $2_{9}$ & $2_{10}$ & $2_{12}$ & $2_{15}$ & $3_{4}$ & $3_{5}$ & $3_{7}$ & $3_{9}$ & $3_{13}$ & $3_{15}$ & $3_{18}$ \\
 & $3_{19}$ & $4_{2}$ & $4_{4}$ & $4_{6}$ & $4_{9}$ &  &  &  &  &  &  &  &  &  & \\
\hline
$6_1$ & $1_{2}$ & $1_{3}$ & $1_{5}$ & $1_{6}$ & $2_{2}$ & $2_{3}$ & $2_{5}$ & $2_{7}$ & $2_{8}$ & $2_{9}$ & $2_{10}$ & $2_{11}$ & $2_{12}$ & $2_{14}$ & $2_{15}$ \\
 & $3_{4}$ & $3_{5}$ & $3_{6}$ & $3_{7}$ & $3_{8}$ & $3_{9}$ & $3_{10}$ & $3_{13}$ & $3_{14}$ & $3_{15}$ & $3_{16}$ & $3_{17}$ & $3_{18}$ & $3_{19}$ & $4_{2}$ \\
 & $4_{3}$ & $4_{4}$ & $4_{5}$ & $4_{6}$ & $4_{8}$ & $4_{9}$ &  &  &  &  &  &  &  &  & \\
\hline
\end{tabular}

%% file: Ml-MD-normal.tex
\begin{tabular}{|c|c|c|c|c|c|c|c|}
\hline
$(M_\ell, M_D)$ & $n$ & $m_0^\mathrm{min}\,\text{[eV]}$ & $m_0^\mathrm{max}\,\text{[eV]}$ & texture predicts \\
\hline
$3_{2}-7_{1}$ & $9$ & $0.00 \times 10^{0}$ & $0.00 \times 10^{0}$ & $m_0$ \\
$3_{2}-7_{3}$ & $9$ & $0.00 \times 10^{0}$ & $0.00 \times 10^{0}$ & $m_0$ \\
$4_{1}-6_{1}$ & $9$ & $0.00 \times 10^{0}$ & $0.00 \times 10^{0}$ & $m_0$ \\
$4_{1}-6_{3}$ & $9$ & $0.00 \times 10^{0}$ & $0.00 \times 10^{0}$ & $m_0$ \\
$4_{1}-6_{4}$ & $9$ & $0.00 \times 10^{0}$ & $0.00 \times 10^{0}$ & $m_0$ \\
$4_{1}-6_{5}$ & $9$ & $0.00 \times 10^{0}$ & $0.00 \times 10^{0}$ & $m_0$ \\
$4_{1}-6_{6}$ & $9$ & $0.00 \times 10^{0}$ & $0.00 \times 10^{0}$ & $m_0$ \\
$4_{2}-6_{1}$ & $9$ & $0.00 \times 10^{0}$ & $0.00 \times 10^{0}$ & $m_0$ \\
$4_{2}-6_{2}$ & $9$ & $0.00 \times 10^{0}$ & $0.00 \times 10^{0}$ & $m_0$ \\
$4_{2}-6_{3}$ & $9$ & $0.00 \times 10^{0}$ & $0.00 \times 10^{0}$ & $m_0$ \\
$4_{2}-6_{7}$ & $9$ & $0.00 \times 10^{0}$ & $0.00 \times 10^{0}$ & $m_0$ \\
$4_{2}-6_{8}$ & $9$ & $0.00 \times 10^{0}$ & $0.00 \times 10^{0}$ & $m_0$ \\
$4_{3}-6_{1}$ & $9$ & $0.00 \times 10^{0}$ & $0.00 \times 10^{0}$ & $m_0$ \\
$4_{3}-6_{2}$ & $9$ & $0.00 \times 10^{0}$ & $0.00 \times 10^{0}$ & $m_0$ \\
$4_{3}-6_{3}$ & $9$ & $0.00 \times 10^{0}$ & $0.00 \times 10^{0}$ & $m_0$ \\
$4_{3}-6_{4}$ & $9$ & $0.00 \times 10^{0}$ & $0.00 \times 10^{0}$ & $m_0$ \\
$4_{3}-6_{5}$ & $9$ & $0.00 \times 10^{0}$ & $0.00 \times 10^{0}$ & $m_0$ \\
$4_{3}-6_{6}$ & $9$ & $0.00 \times 10^{0}$ & $0.00 \times 10^{0}$ & $m_0$ \\
$4_{3}-6_{7}$ & $9$ & $0.00 \times 10^{0}$ & $0.00 \times 10^{0}$ & $m_0$ \\
$4_{3}-6_{8}$ & $9$ & $0.00 \times 10^{0}$ & $0.00 \times 10^{0}$ & $m_0$ \\
$4_{3}-6_{9}$ & $9$ & $0.00 \times 10^{0}$ & $0.00 \times 10^{0}$ & $m_0$ \\
$5_{1}-5_{1}$ & $9$ & $0.00 \times 10^{0}$ & $0.00 \times 10^{0}$ & $m_0$ \\
$5_{1}-5_{4}$ & $9$ & $0.00 \times 10^{0}$ & $0.00 \times 10^{0}$ & $m_0$ \\
$5_{1}-5_{5}$ & $9$ & $0.00 \times 10^{0}$ & $0.00 \times 10^{0}$ & $m_0$ \\
$5_{1}-5_{6}$ & $9$ & $0.00 \times 10^{0}$ & $0.00 \times 10^{0}$ & $m_0$ \\
$5_{1}-5_{8}$ & $9$ & $0.00 \times 10^{0}$ & $0.00 \times 10^{0}$ & $m_0$ \\
$6_{1}-3_{12}$ & $10$ & $1.69 \times 10^{-2}$ & $3.32 \times 10^{-1}$ & --- \\
$6_{1}-4_{1}$ & $9$ & $0.00 \times 10^{0}$ & $0.00 \times 10^{0}$ & $m_0$ \\
$6_{1}-4_{5}$ & $8$ & $1.15 \times 10^{-2}$ & $1.65 \times 10^{-2}$ & $m_0$ \\
\hline
\end{tabular}

%% file: Ml-MD-inverted.tex
\begin{tabular}{|c|c|c|c|c|c|c|c|}
\hline
$(M_\ell, M_D)$ & $n$ & $m_0^\mathrm{min}\,\text{[eV]}$ & $m_0^\mathrm{max}\,\text{[eV]}$ & texture predicts \\
\hline
$3_{2}-7_{1}$ & $9$ & $0.00 \times 10^{0}$ & $0.00 \times 10^{0}$ & $m_0$ \\
$3_{2}-7_{3}$ & $9$ & $0.00 \times 10^{0}$ & $0.00 \times 10^{0}$ & $m_0$ \\
$4_{1}-6_{1}$ & $9$ & $0.00 \times 10^{0}$ & $0.00 \times 10^{0}$ & $m_0$ \\
$4_{1}-6_{3}$ & $9$ & $0.00 \times 10^{0}$ & $0.00 \times 10^{0}$ & $m_0$ \\
$4_{1}-6_{4}$ & $9$ & $0.00 \times 10^{0}$ & $0.00 \times 10^{0}$ & $m_0$ \\
$4_{1}-6_{5}$ & $9$ & $0.00 \times 10^{0}$ & $0.00 \times 10^{0}$ & $m_0$ \\
$4_{1}-6_{6}$ & $9$ & $0.00 \times 10^{0}$ & $0.00 \times 10^{0}$ & $m_0$ \\
$4_{2}-6_{1}$ & $9$ & $0.00 \times 10^{0}$ & $0.00 \times 10^{0}$ & $m_0$ \\
$4_{2}-6_{2}$ & $9$ & $0.00 \times 10^{0}$ & $0.00 \times 10^{0}$ & $m_0$ \\
$4_{2}-6_{3}$ & $9$ & $0.00 \times 10^{0}$ & $0.00 \times 10^{0}$ & $m_0$ \\
$4_{2}-6_{7}$ & $9$ & $0.00 \times 10^{0}$ & $0.00 \times 10^{0}$ & $m_0$ \\
$4_{2}-6_{8}$ & $9$ & $0.00 \times 10^{0}$ & $0.00 \times 10^{0}$ & $m_0$ \\
$4_{3}-6_{1}$ & $9$ & $0.00 \times 10^{0}$ & $0.00 \times 10^{0}$ & $m_0$ \\
$4_{3}-6_{2}$ & $9$ & $0.00 \times 10^{0}$ & $0.00 \times 10^{0}$ & $m_0$ \\
$4_{3}-6_{3}$ & $9$ & $0.00 \times 10^{0}$ & $0.00 \times 10^{0}$ & $m_0$ \\
$4_{3}-6_{4}$ & $9$ & $0.00 \times 10^{0}$ & $0.00 \times 10^{0}$ & $m_0$ \\
$4_{3}-6_{5}$ & $9$ & $0.00 \times 10^{0}$ & $0.00 \times 10^{0}$ & $m_0$ \\
$4_{3}-6_{6}$ & $9$ & $0.00 \times 10^{0}$ & $0.00 \times 10^{0}$ & $m_0$ \\
$4_{3}-6_{7}$ & $9$ & $0.00 \times 10^{0}$ & $0.00 \times 10^{0}$ & $m_0$ \\
$4_{3}-6_{8}$ & $9$ & $0.00 \times 10^{0}$ & $0.00 \times 10^{0}$ & $m_0$ \\
$4_{3}-6_{9}$ & $9$ & $0.00 \times 10^{0}$ & $0.00 \times 10^{0}$ & $m_0$ \\
$5_{1}-5_{1}$ & $9$ & $0.00 \times 10^{0}$ & $0.00 \times 10^{0}$ & $m_0$ \\
$5_{1}-5_{4}$ & $9$ & $0.00 \times 10^{0}$ & $0.00 \times 10^{0}$ & $m_0$ \\
$5_{1}-5_{5}$ & $9$ & $0.00 \times 10^{0}$ & $0.00 \times 10^{0}$ & $m_0$ \\
$5_{1}-5_{6}$ & $9$ & $0.00 \times 10^{0}$ & $0.00 \times 10^{0}$ & $m_0$ \\
$5_{1}-5_{8}$ & $9$ & $0.00 \times 10^{0}$ & $0.00 \times 10^{0}$ & $m_0$ \\
$6_{1}-3_{12}$ & $10$ & $< 10^{-3}$ & $3.31 \times 10^{-1}$ & --- \\
$6_{1}-4_{1}$ & $9$ & $0.00 \times 10^{0}$ & $0.00 \times 10^{0}$ & $m_0$ \\
\hline
\end{tabular}

%% file: Ml-MnuL-normal.tex
\begin{tabular}{|c|c|c|c|c|c|c|c|c|c|}
\hline
$(M_\ell, M_L)$ & $n$ & $\chi^2_\mathrm{min}$ & $m_0^\mathrm{min}\,\text{[eV]}$ & $m_0^\mathrm{max}\,\text{[eV]}$ & $\delta^\mathrm{min}$ & $\delta^\mathrm{max}$ &
$m_{\beta\beta}^\mathrm{min}\,\text{[eV]}$ & $m_{\beta\beta}^\mathrm{max}\,\text{[eV]}$ & texture predicts\\
\hline
$3_{2}-4_{7}$ & $10$ & $<10^{-4}$ & $0.00 \times 10^{0}$ & $0.00 \times 10^{0}$ & $<10^{-2}$ & $6.27 \times 10^{0}$ & $1.34 \times 10^{-3}$ & $3.86 \times 10^{-3}$ & $m_0$, $m_{\beta\beta}$ \\
$3_{2}-4_{9}$ & $10$ & $<10^{-4}$ & $0.00 \times 10^{0}$ & $0.00 \times 10^{0}$ & $<10^{-2}$ & $6.28 \times 10^{0}$ & $1.34 \times 10^{-3}$ & $3.86 \times 10^{-3}$ & $m_0$, $m_{\beta\beta}$ \\
$4_{1}-3_{3}$ & $10$ & $<10^{-4}$ & $1.17 \times 10^{-3}$ & $3.32 \times 10^{-1}$ & $<10^{-2}$ & $6.28 \times 10^{0}$ & $<10^{-3}$ & $2.14 \times 10^{-1}$ & --- \\
$4_{1}-3_{15}$ & $10$ & $2.63 \times 10^{-2}$ & $4.49 \times 10^{-2}$ & $3.32 \times 10^{-1}$ & $2.92 \times 10^{0}$ & $3.37 \times 10^{0}$ & $1.45 \times 10^{-2}$ & $1.34 \times 10^{-1}$ & $\delta$ \\
$4_{1}-4_{1}$ & $8$ & $2.99 \times 10^{1}$ & $0.00 \times 10^{0}$ & $0.00 \times 10^{0}$ & $3.07 \times 10^{0}$ & $3.21 \times 10^{0}$ & $1.57 \times 10^{-3}$ & $1.69 \times 10^{-3}$ & $m_0$, $\delta$, $m_{\beta\beta}$ \\
$4_{1}-4_{4}$ & $8$ & $2.80 \times 10^{1}$ & $0.00 \times 10^{0}$ & $0.00 \times 10^{0}$ & $3.07 \times 10^{0}$ & $3.21 \times 10^{0}$ & $1.56 \times 10^{-3}$ & $1.69 \times 10^{-3}$ & $m_0$, $\delta$, $m_{\beta\beta}$ \\
$4_{2}-3_{2}$ & $10$ & $<10^{-4}$ & $0.00 \times 10^{0}$ & $0.00 \times 10^{0}$ & $<10^{-2}$ & $6.27 \times 10^{0}$ & $1.34 \times 10^{-3}$ & $3.86 \times 10^{-3}$ & $m_0$, $m_{\beta\beta}$ \\
$4_{2}-3_{3}$ & $10$ & $<10^{-4}$ & $<10^{-3}$ & $9.24 \times 10^{-2}$ & $<10^{-2}$ & $6.27 \times 10^{0}$ & $<10^{-3}$ & $6.09 \times 10^{-2}$ & $m_{\beta\beta}$ \\
$4_{2}-3_{6}$ & $10$ & $<10^{-4}$ & $<10^{-3}$ & $9.28 \times 10^{-2}$ & $<10^{-2}$ & $6.27 \times 10^{0}$ & $1.19 \times 10^{-3}$ & $6.10 \times 10^{-2}$ & $m_{\beta\beta}$ \\
$4_{2}-3_{8}$ & $10$ & $<10^{-4}$ & $3.23 \times 10^{-3}$ & $6.10 \times 10^{-3}$ & $<10^{-2}$ & $6.28 \times 10^{0}$ & $<10^{-3}$ & $2.10 \times 10^{-3}$ & $m_0$, $m_{\beta\beta}$ \\
$4_{2}-3_{10}$ & $10$ & $<10^{-4}$ & $0.00 \times 10^{0}$ & $0.00 \times 10^{0}$ & $<10^{-2}$ & $6.28 \times 10^{0}$ & $1.34 \times 10^{-3}$ & $3.86 \times 10^{-3}$ & $m_0$, $m_{\beta\beta}$ \\
$4_{2}-3_{11}$ & $10$ & $<10^{-4}$ & $0.00 \times 10^{0}$ & $0.00 \times 10^{0}$ & $<10^{-2}$ & $6.28 \times 10^{0}$ & $1.34 \times 10^{-3}$ & $3.86 \times 10^{-3}$ & $m_0$, $m_{\beta\beta}$ \\
$4_{2}-3_{12}$ & $10$ & $<10^{-4}$ & $1.71 \times 10^{-3}$ & $4.44 \times 10^{-3}$ & $<10^{-2}$ & $6.28 \times 10^{0}$ & $1.16 \times 10^{-3}$ & $2.71 \times 10^{-3}$ & $m_0$, $m_{\beta\beta}$ \\
$4_{2}-3_{14}$ & $10$ & $<10^{-4}$ & $1.71 \times 10^{-3}$ & $4.44 \times 10^{-3}$ & $<10^{-2}$ & $6.28 \times 10^{0}$ & $1.16 \times 10^{-3}$ & $2.71 \times 10^{-3}$ & $m_0$, $m_{\beta\beta}$ \\
$4_{2}-3_{16}$ & $10$ & $<10^{-4}$ & $0.00 \times 10^{0}$ & $0.00 \times 10^{0}$ & $<10^{-2}$ & $6.28 \times 10^{0}$ & $1.34 \times 10^{-3}$ & $3.86 \times 10^{-3}$ & $m_0$, $m_{\beta\beta}$ \\
$4_{3}-3_{4}$ & $10$ & $<10^{-4}$ & $1.17 \times 10^{-3}$ & $3.32 \times 10^{-1}$ & $<10^{-2}$ & $6.28 \times 10^{0}$ & $<10^{-3}$ & $2.14 \times 10^{-1}$ & --- \\
$4_{3}-3_{9}$ & $10$ & $<10^{-4}$ & $1.17 \times 10^{-3}$ & $3.32 \times 10^{-1}$ & $<10^{-2}$ & $6.28 \times 10^{0}$ & $<10^{-3}$ & $2.14 \times 10^{-1}$ & --- \\
$4_{3}-3_{14}$ & $10$ & $2.63 \times 10^{-2}$ & $4.49 \times 10^{-2}$ & $3.32 \times 10^{-1}$ & $2.92 \times 10^{0}$ & $3.37 \times 10^{0}$ & $1.45 \times 10^{-2}$ & $1.34 \times 10^{-1}$ & $\delta$ \\
$4_{3}-3_{18}$ & $10$ & $2.63 \times 10^{-2}$ & $4.49 \times 10^{-2}$ & $3.32 \times 10^{-1}$ & $2.92 \times 10^{0}$ & $3.37 \times 10^{0}$ & $1.45 \times 10^{-2}$ & $1.34 \times 10^{-1}$ & $\delta$ \\
$4_{3}-4_{2}$ & $8$ & $2.99 \times 10^{1}$ & $0.00 \times 10^{0}$ & $0.00 \times 10^{0}$ & $3.07 \times 10^{0}$ & $3.21 \times 10^{0}$ & $1.57 \times 10^{-3}$ & $1.69 \times 10^{-3}$ & $m_0$, $\delta$, $m_{\beta\beta}$ \\
$4_{3}-4_{3}$ & $8$ & $2.80 \times 10^{1}$ & $0.00 \times 10^{0}$ & $0.00 \times 10^{0}$ & $3.07 \times 10^{0}$ & $3.21 \times 10^{0}$ & $1.56 \times 10^{-3}$ & $1.69 \times 10^{-3}$ & $m_0$, $\delta$, $m_{\beta\beta}$ \\
$4_{3}-4_{4}$ & $8$ & $2.99 \times 10^{1}$ & $0.00 \times 10^{0}$ & $0.00 \times 10^{0}$ & $3.07 \times 10^{0}$ & $3.21 \times 10^{0}$ & $1.57 \times 10^{-3}$ & $1.69 \times 10^{-3}$ & $m_0$, $\delta$, $m_{\beta\beta}$ \\
$4_{3}-4_{5}$ & $8$ & $2.80 \times 10^{1}$ & $0.00 \times 10^{0}$ & $0.00 \times 10^{0}$ & $3.07 \times 10^{0}$ & $3.21 \times 10^{0}$ & $1.56 \times 10^{-3}$ & $1.69 \times 10^{-3}$ & $m_0$, $\delta$, $m_{\beta\beta}$ \\
$5_{1}-3_{1}$ & $8$ & $3.87 \times 10^{0}$ & $2.76 \times 10^{-2}$ & $2.79 \times 10^{-2}$ & $2.11 \times 10^{0}$ & $4.18 \times 10^{0}$ & $2.60 \times 10^{-2}$ & $2.78 \times 10^{-2}$ & $m_0$, $m_{\beta\beta}$ \\
$6_{1}-2_{1}$ & $8$ & $2.25 \times 10^{1}$ & $1.52 \times 10^{-1}$ & $3.32 \times 10^{-1}$ & $<10^{-2}$ & $6.28 \times 10^{0}$ & $1.44 \times 10^{-1}$ & $3.29 \times 10^{-1}$ & --- \\
$6_{1}-2_{4}$ & $8$ & $<10^{-4}$ & $4.00 \times 10^{-3}$ & $4.21 \times 10^{-2}$ & $1.32 \times 10^{0}$ & $4.96 \times 10^{0}$ & $0.00 \times 10^{0}$ & $4.29 \times 10^{-2}$ & $m_0$, $m_{\beta\beta}$ \\
$6_{1}-2_{6}$ & $8$ & $<10^{-4}$ & $3.52 \times 10^{-2}$ & $3.91 \times 10^{-2}$ & $1.64 \times 10^{0}$ & $4.64 \times 10^{0}$ & $3.58 \times 10^{-2}$ & $3.97 \times 10^{-2}$ & $m_0$, $m_{\beta\beta}$ \\
\hline
\end{tabular}

%% file: Ml-MnuL-inverted.tex
\begin{tabular}{|c|c|c|c|c|c|c|c|c|c|}
\hline
$(M_\ell, M_L)$ & $n$ & $\chi^2_\mathrm{min}$ & $m_0^\mathrm{min}\,\text{[eV]}$ & $m_0^\mathrm{max}\,\text{[eV]}$ & $\delta^\mathrm{min}$ & $\delta^\mathrm{max}$ &
$m_{\beta\beta}^\mathrm{min}\,\text{[eV]}$ & $m_{\beta\beta}^\mathrm{max}\,\text{[eV]}$ & texture predicts\\
\hline
$3_{2}-4_{7}$ & $10$ & $<10^{-4}$ & $0.00 \times 10^{0}$ & $0.00 \times 10^{0}$ & $<10^{-2}$ & $6.28 \times 10^{0}$ & $1.79 \times 10^{-2}$ & $4.88 \times 10^{-2}$ & $m_0$, $m_{\beta\beta}$ \\
$3_{2}-4_{9}$ & $10$ & $<10^{-4}$ & $0.00 \times 10^{0}$ & $0.00 \times 10^{0}$ & $<10^{-2}$ & $6.27 \times 10^{0}$ & $1.79 \times 10^{-2}$ & $4.88 \times 10^{-2}$ & $m_0$, $m_{\beta\beta}$ \\
$4_{1}-4_{1}$ & $8$ & $2.43 \times 10^{-3}$ & $0.00 \times 10^{0}$ & $0.00 \times 10^{0}$ & $2.87 \times 10^{0}$ & $3.41 \times 10^{0}$ & $1.80 \times 10^{-2}$ & $1.87 \times 10^{-2}$ & $m_0$, $\delta$, $m_{\beta\beta}$ \\
$4_{1}-4_{2}$ & $8$ & $1.41 \times 10^{-1}$ & $0.00 \times 10^{0}$ & $0.00 \times 10^{0}$ & $2.97 \times 10^{0}$ & $3.31 \times 10^{0}$ & $1.76 \times 10^{-2}$ & $1.82 \times 10^{-2}$ & $m_0$, $\delta$, $m_{\beta\beta}$ \\
$4_{1}-4_{4}$ & $8$ & $2.75 \times 10^{-2}$ & $0.00 \times 10^{0}$ & $0.00 \times 10^{0}$ & $2.92 \times 10^{0}$ & $3.37 \times 10^{0}$ & $1.79 \times 10^{-2}$ & $1.85 \times 10^{-2}$ & $m_0$, $\delta$, $m_{\beta\beta}$ \\
$4_{2}-3_{11}$ & $10$ & $<10^{-4}$ & $0.00 \times 10^{0}$ & $0.00 \times 10^{0}$ & $<10^{-2}$ & $6.28 \times 10^{0}$ & $1.79 \times 10^{-2}$ & $4.88 \times 10^{-2}$ & $m_0$, $m_{\beta\beta}$ \\
$4_{2}-3_{14}$ & $10$ & $<10^{-4}$ & $1.73 \times 10^{-3}$ & $4.26 \times 10^{-3}$ & $<10^{-2}$ & $6.27 \times 10^{0}$ & $1.78 \times 10^{-2}$ & $4.89 \times 10^{-2}$ & $m_0$, $m_{\beta\beta}$ \\
$4_{2}-3_{16}$ & $10$ & $<10^{-4}$ & $0.00 \times 10^{0}$ & $0.00 \times 10^{0}$ & $<10^{-2}$ & $6.27 \times 10^{0}$ & $1.79 \times 10^{-2}$ & $4.88 \times 10^{-2}$ & $m_0$, $m_{\beta\beta}$ \\
$4_{2}-3_{17}$ & $10$ & $<10^{-4}$ & $1.73 \times 10^{-3}$ & $4.26 \times 10^{-3}$ & $<10^{-2}$ & $6.27 \times 10^{0}$ & $1.78 \times 10^{-2}$ & $4.89 \times 10^{-2}$ & $m_0$, $m_{\beta\beta}$ \\
$4_{3}-4_{1}$ & $8$ & $1.41 \times 10^{-1}$ & $0.00 \times 10^{0}$ & $0.00 \times 10^{0}$ & $2.97 \times 10^{0}$ & $3.31 \times 10^{0}$ & $1.76 \times 10^{-2}$ & $1.82 \times 10^{-2}$ & $m_0$, $\delta$, $m_{\beta\beta}$ \\
$4_{3}-4_{2}$ & $8$ & $2.43 \times 10^{-3}$ & $0.00 \times 10^{0}$ & $0.00 \times 10^{0}$ & $2.87 \times 10^{0}$ & $3.41 \times 10^{0}$ & $1.80 \times 10^{-2}$ & $1.87 \times 10^{-2}$ & $m_0$, $\delta$, $m_{\beta\beta}$ \\
$4_{3}-4_{3}$ & $8$ & $2.75 \times 10^{-2}$ & $0.00 \times 10^{0}$ & $0.00 \times 10^{0}$ & $2.92 \times 10^{0}$ & $3.37 \times 10^{0}$ & $1.79 \times 10^{-2}$ & $1.85 \times 10^{-2}$ & $m_0$, $\delta$, $m_{\beta\beta}$ \\
$4_{3}-4_{4}$ & $8$ & $2.43 \times 10^{-3}$ & $0.00 \times 10^{0}$ & $0.00 \times 10^{0}$ & $2.87 \times 10^{0}$ & $3.41 \times 10^{0}$ & $1.80 \times 10^{-2}$ & $1.87 \times 10^{-2}$ & $m_0$, $\delta$, $m_{\beta\beta}$ \\
$4_{3}-4_{5}$ & $8$ & $2.75 \times 10^{-2}$ & $0.00 \times 10^{0}$ & $0.00 \times 10^{0}$ & $2.92 \times 10^{0}$ & $3.37 \times 10^{0}$ & $1.79 \times 10^{-2}$ & $1.85 \times 10^{-2}$ & $m_0$, $\delta$, $m_{\beta\beta}$ \\
$4_{3}-4_{6}$ & $8$ & $1.41 \times 10^{-1}$ & $0.00 \times 10^{0}$ & $0.00 \times 10^{0}$ & $2.97 \times 10^{0}$ & $3.31 \times 10^{0}$ & $1.76 \times 10^{-2}$ & $1.82 \times 10^{-2}$ & $m_0$, $\delta$, $m_{\beta\beta}$ \\
$5_{1}-3_{1}$ & $8$ & $<10^{-4}$ & $<10^{-3}$ & $<10^{-3}$ & $2.83 \times 10^{0}$ & $3.46 \times 10^{0}$ & $1.81 \times 10^{-2}$ & $1.89 \times 10^{-2}$ & $m_0$, $\delta$, $m_{\beta\beta}$ \\
$5_{1}-3_{2}$ & $8$ & $1.80 \times 10^{-1}$ & $0.00 \times 10^{0}$ & $0.00 \times 10^{0}$ & $2.98 \times 10^{0}$ & $3.31 \times 10^{0}$ & $1.75 \times 10^{-2}$ & $1.81 \times 10^{-2}$ & $m_0$, $\delta$, $m_{\beta\beta}$ \\
$5_{1}-3_{3}$ & $8$ & $<10^{-4}$ & $1.05 \times 10^{-3}$ & $1.14 \times 10^{-3}$ & $2.58 \times 10^{0}$ & $3.71 \times 10^{0}$ & $1.95 \times 10^{-2}$ & $2.04 \times 10^{-2}$ & $m_0$, $\delta$, $m_{\beta\beta}$ \\
$5_{1}-3_{6}$ & $8$ & $5.64 \times 10^{-1}$ & $<10^{-3}$ & $<10^{-3}$ & $3.01 \times 10^{0}$ & $3.27 \times 10^{0}$ & $1.70 \times 10^{-2}$ & $1.76 \times 10^{-2}$ & $m_0$, $\delta$, $m_{\beta\beta}$ \\
$5_{1}-3_{7}$ & $8$ & $<10^{-4}$ & $1.05 \times 10^{-3}$ & $1.14 \times 10^{-3}$ & $2.58 \times 10^{0}$ & $3.71 \times 10^{0}$ & $1.95 \times 10^{-2}$ & $2.04 \times 10^{-2}$ & $m_0$, $\delta$, $m_{\beta\beta}$ \\
$5_{1}-3_{8}$ & $8$ & $5.64 \times 10^{-1}$ & $<10^{-3}$ & $<10^{-3}$ & $3.01 \times 10^{0}$ & $3.27 \times 10^{0}$ & $1.70 \times 10^{-2}$ & $1.76 \times 10^{-2}$ & $m_0$, $\delta$, $m_{\beta\beta}$ \\
$5_{1}-3_{10}$ & $8$ & $1.80 \times 10^{-1}$ & $0.00 \times 10^{0}$ & $0.00 \times 10^{0}$ & $2.98 \times 10^{0}$ & $3.31 \times 10^{0}$ & $1.75 \times 10^{-2}$ & $1.81 \times 10^{-2}$ & $m_0$, $\delta$, $m_{\beta\beta}$ \\
$6_{1}-2_{1}$ & $8$ & $<10^{-4}$ & $3.31 \times 10^{-2}$ & $4.02 \times 10^{-2}$ & $8.65 \times 10^{-1}$ & $5.42 \times 10^{0}$ & $3.26 \times 10^{-2}$ & $3.99 \times 10^{-2}$ & $\mathrm{sin}^2\theta_{23}$, $m_0$, $m_{\beta\beta}$ \\
$6_{1}-2_{4}$ & $8$ & $<10^{-4}$ & $4.10 \times 10^{-2}$ & $4.72 \times 10^{-2}$ & $1.55 \times 10^{0}$ & $4.73 \times 10^{0}$ & $6.37 \times 10^{-2}$ & $6.80 \times 10^{-2}$ & $m_0$, $m_{\beta\beta}$ \\
$6_{1}-2_{6}$ & $8$ & $<10^{-4}$ & $3.80 \times 10^{-2}$ & $4.39 \times 10^{-2}$ & $1.57 \times 10^{0}$ & $4.71 \times 10^{0}$ & $6.19 \times 10^{-2}$ & $6.58 \times 10^{-2}$ & $m_0$, $m_{\beta\beta}$ \\
\hline
\end{tabular}